\documentclass[aip,jmp,amsmath,amssymb,preprint,
]{revtex4-1}

\usepackage{graphicx}
\usepackage{dcolumn}
\usepackage{bm}
\usepackage[colorlinks=true,linkcolor=blue]{hyperref}%

\begin{document}

\preprint{AIP/123-QED}

\title[R. Arun, P. Sabareesan and M. Daniel]{Dynamics of Current and Field Driven Domain Wall Motion under the Influence of Transverse Magnetic Field}
\author{R. Arun}%
\email{arunbdu@gmail.com}
\affiliation{Centre for Nonlinear Dynamics, School of Physics, Bharathidasan University, Tiruchirapalli - 620 024, Tamilnadu, India.
}%
\author{P. Sabareesan}
\email{sabaribard@gmail.com}
\affiliation{
Centre for Nonlinear Science and Engineering, School of Electrical and Electronics Engineering, SASTRA University, Thanjavur-613 401, Tamilnadu, India.
}%

\author{M. Daniel}
\email{danielcnld@gmail.com}
\affiliation{Centre for Nonlinear Dynamics, School of Physics, Bharathidasan University, Tiruchirapalli - 620 024, Tamilnadu, India.
}%
\begin{abstract}{
The dynamics of transverse Neel domain wall in a ferromagnetic nanostrip in the presence of driving field, current and transverse magnetic field is investigated by the Landau-Lifshitz-Gilbert(LLG) equation with the adiabatic and non-adiabatic spin-transfer torques both analytically and numerically. The analytical expressions for the velocity, width, excitation angle and displacement for the domain wall are obtained by using small angle approximation along with Walkers trial function. The results show that the initial velocity of the domain wall can be controlled by the adiabatic spin-transfer torque and the saturated velocity can be controlled by the non-adiabatic spin-transfer torque and driving field.  The large increase in the saturated velocity of the domain wall driven by current and field due to the transverse magnetic field is identified through the presence of driving field.  There is no impact in the saturated velocity of the domain wall driven by current from the transverse magnetic field.  For the domain wall driven by the current in the presence of the transverse magnetic field, the saturated velocity remains constant. The transverse magnetic field along with current and  driving field is more advantageous that the transverse magnetic field along with current for increasing the saturated velocity of the domain wall.  The numerical results showed that the saturated velocity is increased by the transverse magnetic field with the irrespective of the directions of the driving field and current further it is higher and lower when the directions of driving field and current are antiparallel and parallel respectively. The obtained analytical solutions are closely coincided with the computed numerical results.}

\end{abstract}

\maketitle

\section{Introduction}
Domain wall motion in magnetic nanostructures driven by a magnetic field and current have variety of potential and technological applications including logic devices\cite{Allwood}, atom trapping\cite{Allwood1} and memory storage\cite{Parkin,Parkin1,Grollier,Zhao}. Both the field induced and current induced domain wall dynamics have been extensively studied experimentally\cite{Hayashi,Ono,Atkinson,Nakatani,Beach,Lewis,Thomas,Thomas1,Grollier,Boone} and numerically\cite{Schryer,Yang,Tretiakov,Albert,Lee,Ai,Yan,Emoria,Thiaville}.  The transverse domain wall has higher velocity and less complexity in motion in the presence of  magnetic field\cite{He}.  In the presence of current or field the transverse Neel wall in nanostrip can be moved along the length axis and the average speed of the wall is almost linear with field  or current below the critical value,  the so called Walker limit of field or current respectively. The domain wall moves rigidly with slight excitation and distortion below the Walker limit whereas above the Walker limit, the motion of the domain wall is no more linear and the average speed of the wall reduces suddenly. These have been stuided analytically\cite{Schryer,Mougin,Li}, numerically\cite{Schryer,Li}, experimentally\cite{Beach,Beach1} and also through micromagnetic simulation\cite{Yang,Tretiakov,Thiaville,Djuhana,Fukami,Jang,Khvalkovskiy}. The drastic decrease in the speed of the domain wall is due to the oscillatory behaviour that occurs in the structure of the wall, which depends on the cross sectional area of the strip. When the width and thickness of the strip are large, the domain wall oscillates between transverse \cite{Jun,Klaui} and vortex type domain wall, whereas in the case of the width and thickness are small, the transverse wall rotates about the length axis\cite{Zhang,Li,Mougin,Szambolics}. There are few methods available to increase the speed of the domain wall and the Walker limit: introducing roughness in the strip\cite{Nakatani}, making nanostrip with honey comb structure\cite{Lewis}, applying oscillatory magnetic field\cite{Weerts}, inclusion of perpendicular anisotropy under layer\cite{Lee1} and by applying a transverse magnetic field.

 The recent experimental\cite{Glathe,Glathe1,Richter} and micromagnetic simulation\cite{Kunz,Matthew,Lu} studies on field induced domain wall dynamics show that when a transverse magnetic field is applied parallel to the magnetic moments of the domain wall and it maintains the regular motion of the domain wall even for large applied field. This implies that, the Walker limit is increased and it corresponds to increase the speed of the domain wall. Also, the presence of the transverse magnetic field, increases the width of the domain wall\cite{Kunz,Matthew} and making asymmetry and twisting in the domain wall\cite{Lu}. However, when a transverse magnetic field is applied in a direction antiparallel to the direction of magnetic moments of the domain wall, the speed and width of the domain wall as well as the Walker limit decreases\cite{Kunz,Matthew}.

Analytical and micromagnetic simulation study of current induced Bloch wall\cite{Boulle} motion in perpendicularly magnetized nanostrip and micromagnetic simulation study of Neel wall\cite{Jang} motion in parallely magnetized nanostrip show that there is an increase in the Walker limit as well as the speed of the domain wall. Inspite of the above developments in the case of field induced domain wall motion in the presence of a transverse magnetic field, the current driven domain wall dynamics along with transverse magnetic field is yet to be understood properly, especially for the transverse Neel domain wall. A systematic analytical study on the dynamics of Neel domain wall driven by current and field in the presence of a transverse magnetic field is not available in the literature. 

Motivated by the above, in the present paper, the dynamics of Neel domain wall in a ferromagnetic nanostrip driven by both the current and field in the presence of a transverse magnetic field are extensively studied through analytically and numerically.  The analytical expression for the excitation angle of the magnetization inside the domain wall, the velocity, the width and the displacement of the domain wall are obtained. The paper is organised as follows: In Section II, we present the model and the governing equation of motion for the dynamics of domain wall in a ferromagnetic nanostrip.  In Section III, the dynamical equation is analytically solved using Walker's trial function and the dynamical parameters such as excitation angle, velocity, width and displacement are derived.  The numerical results are obtained by solving the dynamical equation using Runge-Kutta-4 method and the results are compared with the analytical results in Section IV and the effects of current, driving field and transverse magnetic field on the dynamics of domain wall are discussed.  The influence of the transverse magnetic field in saturated velocity is also observed.  Finally, the results are concluded in Section V.
\section{Dynamical Model for Neel type domain wall motion}
\begin{figure}[!hbtp]
\centering\includegraphics[angle=0,width=0.6\linewidth]{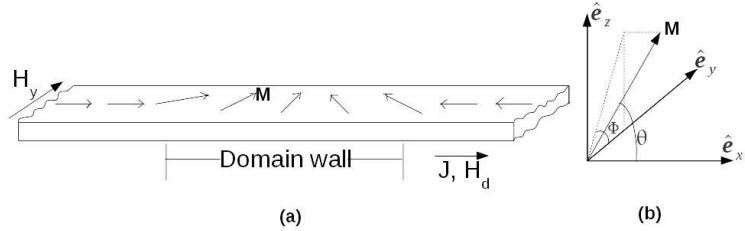}
\caption{ (a) A schematic representation of ferromagnetic nanostrip which is taken as our model with its easy axis of magnetization along x-direction.  The current density ${\bf J}$ and external magnetic field ${\bf H}_{d}$ can be applied in both the directions of x-axis and the velocity of domain wall depends upon the directions of field and current.  The transverse magnetic field ${\bf H}_y$ is applied along positive y-direction.}
\end{figure}
 Consider an infinitely long anisotropic ferromagnetic nanostrip with an easy axis of magnetization along x-direction as shown in FIG.1. In the nanostrip, the uniformly magnetized left domain along +x direction and the right domains along -x direction are separated by a domain wall, where the magnetization can be represented by the vector ${\bf M}(x,t)$. In FIG.1(a), ${\bf H}_{d}$ and ${\bf J}$ represent the driving field and current density applied along x-direction.  The transverse magnetic field ${\bf H}_{y}$ is applied along the positive y-direction. The arrows indicate the direction of magnetization along the nanostip. In FIG.1(b), ${\bf \hat{e}}_x$, ${\bf \hat{e}}_y$ and ${\bf \hat{e}}_z$ represent the unit vectors along x, y and z-directions respectively.  It is assumed that the magnetization of the nanostrip varies only along the x-direction and the initial profile of the magnetization of the model appears as shown in FIG.1(a).  The angles $\theta(x,t)$  and $\Phi(x,t)$ represent the orientation of magnetization with reference to the positive x-direction and its projection in the yz-plane making with the positive y-direction respectively.  Physically the angle $\theta$ represents the spatial variation of the direction of the magnetization along x-direction and $\Phi$ represents the out-of-plane excitation of magnetization vector in the strip.  The Landau-Lifshitz-Gilbert(LLG) equation that governs the dynamics of the magnetization present in the strip in the presence of an externally applied fields and current is written as\cite{Zhang}
\begin{align}
&\frac {\partial {{\bf M}({x},t)}}{\partial t}  ~=~  -\gamma {\bf M} \times {\bf H}_{eff} + \frac{\alpha }{M_{s}} {\bf M} \times \frac{\partial {\bf M}}{\partial t} -\frac{b}{M_{s}^2} {\bf M} \times \left({\bf M} \times \frac{\partial {\bf M}}{\partial x}\right)- \frac {c}{M_{s}} {\bf M} \times  \frac{\partial {\bf M}}{\partial x}, \label{LLG1}\\
&{\bf M}=(M_x,M_y,M_z); ~|{\bf M}|^2=M_x^2+M_y^2+M_z^2~=~M_s^2. \nonumber
\end{align}
Here, ${\bf M}(x,t)$ represents the magnetization vector, $\gamma$ is the gyromagnetic ratio, $\alpha$ is the Gilbert damping parameter, $M_s(=|{\bf M}|)$ is the saturation magnetization of the magnetic strip and $b=P J \mu_B/ e M_s$, $c=\xi b$ represent the magnitude of adiabatic and non-adiabatic spin-transfer torques respectively.  Where, $P$ is the polarization, $J$ is the magnitude of the current density, $\mu_B$ is the Bohr magneton, $e$ is the charge of the electron and $\xi$ is the non-adiabaticity factor. In Eq.\eqref{LLG1}, ${\bf H}_{eff}$ represents the effective field due to different magnetic contributions including exchange interaction, magnetocrystalline anisotropy, the driving field, the transverse magnetic field and the demagnetization field.   The first term in the right hand side of Eq.\eqref{LLG1} represents the precession of the magnetization about the effective field ${\bf H}_{{eff}}$, that determines the precessional frequency and conserves the magnetic energy. The second term, supports the damping of the magnetization due to dissipation of energy that takes place within the material.  The third term, which represents the adiabatic spin-transfer torque corresponding to the reaction torque on the magnetization produced by the spatial variation of the spin current density\cite{Zhang1}.  The last term  represents the non-adiabatic spin-transfer torque, which corresponds to the reaction torque on magnetization due to the continuous space variation of spatially mistraking spins between conduction electrons and local magnetization.  Adding all the above fields, the total effective field can be written as
\begin{align}
{\bf H}_{eff} = \frac{2A}{M_{s}^2} ~\frac{\partial^2 {\bf M}}{\partial x^2} + \left(\frac{H_{k}}{M_{s}}M_{x}+H_d \right) {\bf e}_x + H_y {\bf e}_y - 4\pi M_{z} {\bf e}_z, \label{Heff1}
\end{align}
where $A$ represents the exchange interaction coefficient and $H_k$ represents the magnetocrystalline anisotropy coefficient.  $H_d$, $H_y$ and 4$\pi M_z$ are the magnitudes of driving field, transverse magnetic field and demagnetization field respectively.

The dynamics of magnetization in the domain wall can be understood by solving Eq.\eqref{LLG1} after substituting Eq.\eqref{Heff1} in it.  While solving the dynamical Eq.\eqref{LLG1}, we also evaluate the dynamical quantities of the domain wall namely velocity, width, excitation of the wall from the plane of the strip and the displacement analytically.
\section{Analytical solutions for the domain wall parameters}
As Eq.\eqref{LLG1} is a highly nontrivial vector nonlinear evolution equation, it may be difficult to solve the same in its present form.  Hence, we rewrite Eq.\eqref{LLG1} in terms of the angles $\theta(x,t)$ and $\Phi(x,t)$ by defining the component of magnetization in the polar form as follows.
\begin{subequations}
\label{spherical}
\begin{align}
M_{x} &= M_{s} \cos\theta(x,t),\label{spherical1}\\
M_{y} &= M_{s} \sin\theta(x,t) \cos\Phi(x,t),\label{spherical2}\\
M_{z} &= M_{s} \sin\theta(x,t) \sin\Phi(x,t).\label{spherical3}
\end{align}
\end{subequations}
On substituting Eqs.\eqref{spherical} in Eq.\eqref{LLG1} along with $c=\xi b$, we obtain the following set of equations for $\theta(x,t)$ and $\Phi(x,t)$.
\begin{subequations}
\label{llg}
\begin{align}
\frac{\partial \theta(x,t)}{\partial t}& + \alpha \sin\theta(x,t) \frac{\partial \Phi(x,t)}{\partial t}  ~=~  \gamma\left[\frac{2 A}{M_s}\left(2 \cos\theta(x,t) \frac{\partial \theta(x,t)}{\partial x} \frac{\partial \Phi(x,t)}{\partial x} + \sin\theta(x,t) \frac{\partial^2 \Phi(x,t)}{\partial x^2} \right)\right.\nonumber\\ 
&\left.- H_y \sin\Phi(x,t)- 2\pi M_s \sin\theta(x,t) \sin2\Phi(x,t)\right] + b \left(\frac{\partial \theta(x,t)}{\partial x} + \xi b \sin\theta(x,t) \frac{\partial \Phi(x,t)}{\partial x}\right), \label{llg1} \\
\alpha\frac{\partial \theta(x,t)}{\partial t}&- \sin\theta(x,t) \frac{\partial \Phi(x,t)}{\partial t} =~ \gamma\left[\frac{2 A}{M_{s}}\left(\frac{\partial^2 \theta(x,t)}{\partial x^2}-\sin \theta(x,t) \cos\theta(x,t)\left(\frac{\partial \Phi(x,t)}{\partial x}\right)^2\right)\right.\notag\\
&\left.- H_{d} \sin\theta(x,t)+  H_y \cos\Phi(x,t) \cos\theta(x,t)-\frac{1}{2}\left[H_k + 4\pi M_s \sin^2\Phi(x,t)\right]\sin2\theta(x,t)\right]  \notag\\
&+ b\left[\xi \frac{\partial \theta(x,t)}{\partial x}- \sin\theta(x,t) \frac{\partial \Phi(x,t)}{\partial x}\right]  . \label{llg2}
\end{align}
\end{subequations}
Eqs.\eqref{llg} describe the dynamics of domain wall in a ferromagnetic nanostrip in the presence of current, driving field and transverse magnetic field in terms of $\theta$ and $\Phi$.  The problem now boils down to solving Eqs.\eqref{llg} for $\theta$, $\Phi$ in order to find the excitation angle, velocity and width of the domain wall.
 The same form of Eqs.\eqref{llg} with and without transverse magnetic field in the absence of current has been studied and solved by Schryer $et~al$\cite{Schryer} and Lu $et~al$\cite{Lu} respectively using the trial functions obtained from the static profiles $\theta(x/W)$ and $\Phi(x)$, which are the spatial variations of $\theta$ and $\Phi$ when the domain wall is at rest and $W$ is the width of the domain wall.  Similarly, Li and Zhang\cite{Li} have studied the dynamics of transverse Neel wall in the presence of driving field and current, without transverse magnetic field, using the trial functions obtained by Schryer and Walker\cite{Schryer}.   In the present paper, the dynamics of transverse Neel wall driven by driving field and current in the presence of transverse magnetic field is studied by solving the Eqs.\eqref{llg} using Schryer and Walker's trial functions, which are given by
\begin{subequations}
\label{trialfunction}
\begin{align}
\theta(x,t) &= \theta\left(\frac{x-X(t)}{W(t)}\right),\label{trialtheta}\\
\Phi(x,t) &= \Phi(x) + \phi(t)~ U\left(\frac{x-X(t)}{W(t)} \right). \label{trialphi}
\end{align}
\end{subequations}
Where $X(t)$ is the position of the center of the domain wall and $U$ is a step function which can be defined as
\begin{subequations}
\label{unit}
\begin{align}
&U=1,~~~~~ \mathrm{when},~~~~~ \left(\frac{x-X(t)}{W(t)}\right)<\pi/2, \label{unit1}\\
&\mathrm{and}\nonumber\\
&U=0,~~~~~ \mathrm{when},~~~~~ \left(\frac{x-X(t)}{W(t)}\right)>\pi/2\label{unit2}.
\end{align}
\end{subequations}
The new assigned function $\phi(t)$ can be defined as the angle $\Phi$ of the magnetization at the center of the domain wall, can be called as excitation angle.  In the forthcoming sections, the trial functions will be obtained after deriving the static profiles of $\theta$ and  $\Phi$.

\subsection{Static profile for the domain wall in the presence of transverse magnetic field}
Following Schryer and Walker's\cite{Schryer} procedure, it is required to find out the trial functions of $\theta$ and $\Phi$ from their static profiles and these trial functions are to be substituted in Eqs.\eqref{llg1} and \eqref{llg2} to determine  the excitation angle, the velocity and the width of the domain wall.  In the present work, the static profiles for $\theta$ and $\Phi$ are derived after the transverse magnetic field is applied when the current and driving field are switched off. When a transverse magnetic field $H_y$ is applied along the positive y-direction, it exerts a torque on the magnetic moments in the strip and changes their direction from the initial equilibrium direction to the new equilibrium direction in the duration of few piccoseconds and as a result of this, the magnetization inside the domains are tilted towards the positive y-direction from the direction of easy axis. Therefore, the value of $\theta$ in the entire strip except at the centre of the domain wall changes to the new orientation in the presence of a transverse magnetic field and correspondingly the width of the domain wall increases. Whereas there is no variation in $\Phi$ and it takes the value of zero for the entire strip even after the transverse magnetic field is switched on, because the transverse magnetic field is applied along the same plane in which the magnetic moments are present.  Eventhough, the direction of magnetization in the domains changes, there is no spatial variation in the direction of magnetization inside the domains and thus the domain wall is static in the presence of a transverse magnetic field. Hence, the static profiles can be found in the presence of a transverse magnetic field and absence of current and driving field.  In the absence of any external field and current, the value of $\theta$ is 0 and $\pi$ in the left and right domains respectively and it varies from 0 to $\pi$ inside the domain wall from left to right. When the transverse magnetic field is switched on, the values of $\theta$ in the left and right domains are changed and they can be represented as $\theta_D$ and $\pi-\theta_D$ respectively.
 
\begin{figure}[!hbtp]
\centering\includegraphics[angle=0,width=0.8\linewidth]{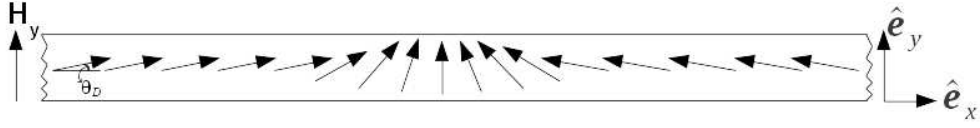}\\
\caption{The spatial variation of magnetization vector of the nanostrip in the presence of a transverse magnetic field $H_y$ applied along the positive y-direction and in the absence of a driving field and current.  The angle $\theta$ varies from $\theta_D$ to $\pi-\theta_D$ from left edge to the right edge along the x-direction and the angle $\Phi$ is zero in the entire strip.}
\end{figure}

Static profile $\theta(x)$ can be computed in the presence of transverse magnetic field once the current and driving field are switched off and the following equation is obtained after substituting $H_d=b=0$ and $\Phi=0$ in Eq.\eqref{llg1}.

\begin{align}
\frac{d^2\theta}{dx^2} - \frac{H_k M_s}{2A}\sin\theta \cos\theta + \frac{H_y M_s}{2A} \cos\theta = 0. \label{staticprofile1}
\end{align}
After multiplying Eq.\eqref{staticprofile1} by $\frac{d\theta}{dx}$, it is integrated with respect to $x$ as
\begin{align}
\left(\frac{d\theta}{dx}\right)^2 - \frac{H_k M_s}{2A}\sin^2\theta + \frac{H_y M_s}{A} \sin\theta = C_1,\label{staticprofile2}
\end{align}
where $C_1$ is the constant of integration and it is obtained  after substituting $\frac{d\theta}{dx}=0$ and $\theta=\theta_D$(for the left domain) in Eq.\eqref{staticprofile2} as
\begin{align}
C_1 = \frac{H_y M_s}{A} \sin\theta_D - \frac{H_k M_s}{2A}\sin^2\theta_D.  \label{staticprofile2a}
\end{align}

The value $\theta_D$ is found from Eq.\eqref{llg2} after substituting $b,H_d=0$ and $\theta=\theta_D, \Phi=0$ corresponding to the left domain as follows:

\begin{align}
\sin\theta_D = \left(\frac{H_y}{H_k}\right),~\theta_D = \sin^{-1}\left(\frac{H_y}{H_k}\right). \label{thetaD2}
\end{align}
$\theta_D$ from Eq.\eqref{thetaD2} gives the angle between the direction of magnetization in the left domain and positive x-direction in the presence of transverse magnetic field.  Knowing $\theta_D$, the angle between the magnetization and positive x-direction in the right domain is calculated as ($\pi-\theta_D$).  From Eq.\eqref{thetaD2} one can understand that when $H_y=H_k$, $\theta_D$ becomes $\pi/2$, which means that all the magnetic moments will be aligned parallel to the direction of the applied transverse magnetic field, which leads to the disappearence of the domain wall.  Therefore one can set the condition for the transverse magnetic field as $H_y<H_k$.  By substituting Eq.\eqref{thetaD2} in Eq.\eqref{staticprofile2a}, $C_1$ is obtained as
\begin{align}
C_1 =  \frac{M_s H_y^2}{2AH_k}. \label{C1}
\end{align}
On substituting the value of $C_1$ from Eq.\eqref{C1} in Eq.\eqref{staticprofile2} one obtains
\begin{align}
\frac{d\theta}{dx} = \pm\sqrt{\frac{H_k M_s}{2A}}\left(\sin\theta-\frac{H_y}{H_k}\right). \label{dthetadx}
\end{align}
The positive sign in Eq.\eqref{dthetadx} represents the variation of $\theta$ from 0 to $\pi$ in the strip along the positive x-direction(head-to-head domain wall) whereas the negative sign is corresponding to the variation from $\pi$ to 0(tail-to-tail domain wall).  In the present model (FIG.1), the head-to-head domain wall configuration is considered to study the domain wall dynamics, and therefore in Eq.\eqref{dthetadx}, only the positive sign survives.
On integrating Eq.\eqref{dthetadx} we obtain
\begin{align}
\frac{1}{\sqrt{1-\left(\frac{H_y}{H_k}\right)^2}} \ln\left\{ \frac{1-\frac{H_y}{H_k}\tan\frac{\theta}{2}-\sqrt{1-\left(\frac{H_y}{H_k}\right)^2}}{1-\frac{H_y}{H_k}\tan\frac{\theta}{2}+\sqrt{1-\left(\frac{H_y}{H_k}\right)^2}}\right\}= \sqrt{\frac{H_k M_s}{2A}} x + C_2, \label{dthetadx3}
\end{align}
where $C_2$ is the constant of integration which can be determined by substituting $\theta=\pi/2$ and $x=0$ in Eq.\eqref{dthetadx3}.  The result reads
\begin{align}
C_2 = \frac{1}{\sqrt{1-\left(\frac{H_y}{H_k}\right)^2}} \ln\left\{\frac{\sqrt{1-\frac{H_y}{H_k}}-\sqrt{1+\frac{H_y}{H_k}}}{\sqrt{1-\frac{H_y}{H_k}}+\sqrt{1+\frac{H_y}{H_k}}} \right\}.\label{dthetadx4}
\end{align}
On substituting $C_2$ in Eq.\eqref{dthetadx3}, one can derive the static profile for $\theta$ as
\begin{align}
\theta(x) = 2~\tan^{-1}\left(\frac{a_1+a_2~\exp\left(\frac{x}{W}\right)}{a_2+a_1~\exp\left(\frac{x}{W}\right)}\right),\label{statictheta}
\end{align}
where, $a_1=\sqrt{1+\frac{H_y}{H_k}}-\sqrt{1-\frac{H_y}{H_k}},~a_2=\sqrt{1+\frac{H_y}{H_k}}+\sqrt{1-\frac{H_y}{H_k}},~W = W_0/\sqrt{1-\left({\frac{H_y}{H_k}}\right)^2}~\mathrm{and}~W_0 = \sqrt{\frac{2A}{H_k M_s}}$.  $W$ and $W_0$ are the width of the domain wall in the presence and absence of transverse magnetic field respectively when current and driving field are switched off. 
The Eq.\eqref{statictheta} represents the spatial variation of the angle between ${\bf M}$ and positive x-direction after the transverse magnetic field is applied while the driving field and current are switched off.
 In the absence of transverse magnetic field ($H_y=0$), Eq.\eqref{statictheta} reduces to $2\tan^{-1}(\exp(x/W_0))$, which agrees with the result obtained by Li and Zhang\cite{Li}. The components of the magnetization in the strip in the absence of current and driving field and in the presence of transverse magnetic field are obtained by substituting the Eq.\eqref{statictheta} and $\Phi=0$ in Eqs.\eqref{spherical}.
\begin{subequations}
\label{M1}
\begin{align}
{M_x} &=  {M_s}\cos\left(2~\tan^{-1}\left[\frac{a_1+a_2~\exp\left(\frac{x}{W}\right)}{a_2+a_1~\exp\left(\frac{x}{W}\right)}\right]\right), \label{Mx1}\\
{M_y} &=  {M_s}\sin\left(2~\tan^{-1}\left[\frac{a_1+a_2~\exp\left(\frac{x}{W}\right)}{a_2+a_1~\exp\left(\frac{x}{W}\right)}\right]\right), \label{My1}\\
{M_z} &= 0. \label{Mz1}
\end{align}
\end{subequations}
\begin{figure}[!hbtp]
\centering\includegraphics[angle=0,width=0.5\linewidth]{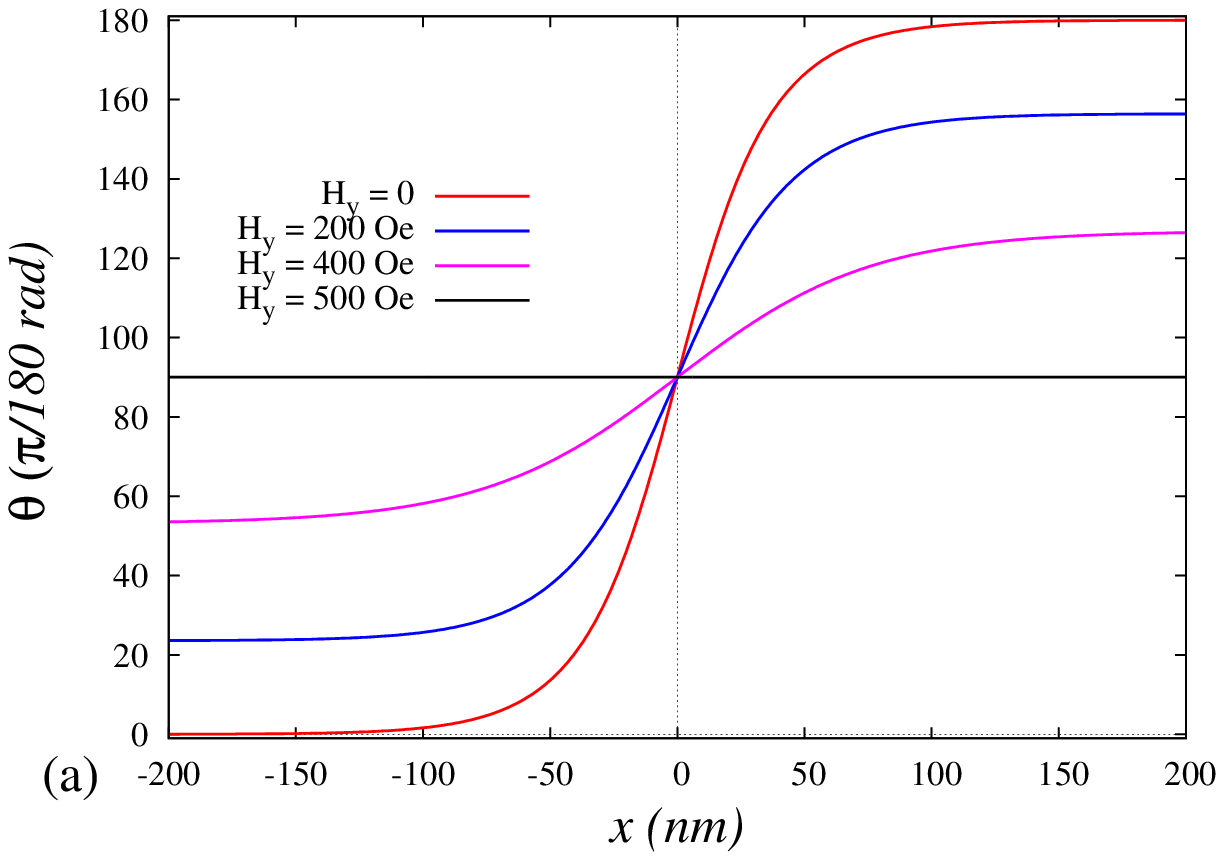}~\includegraphics[angle=0,width=0.5\linewidth]{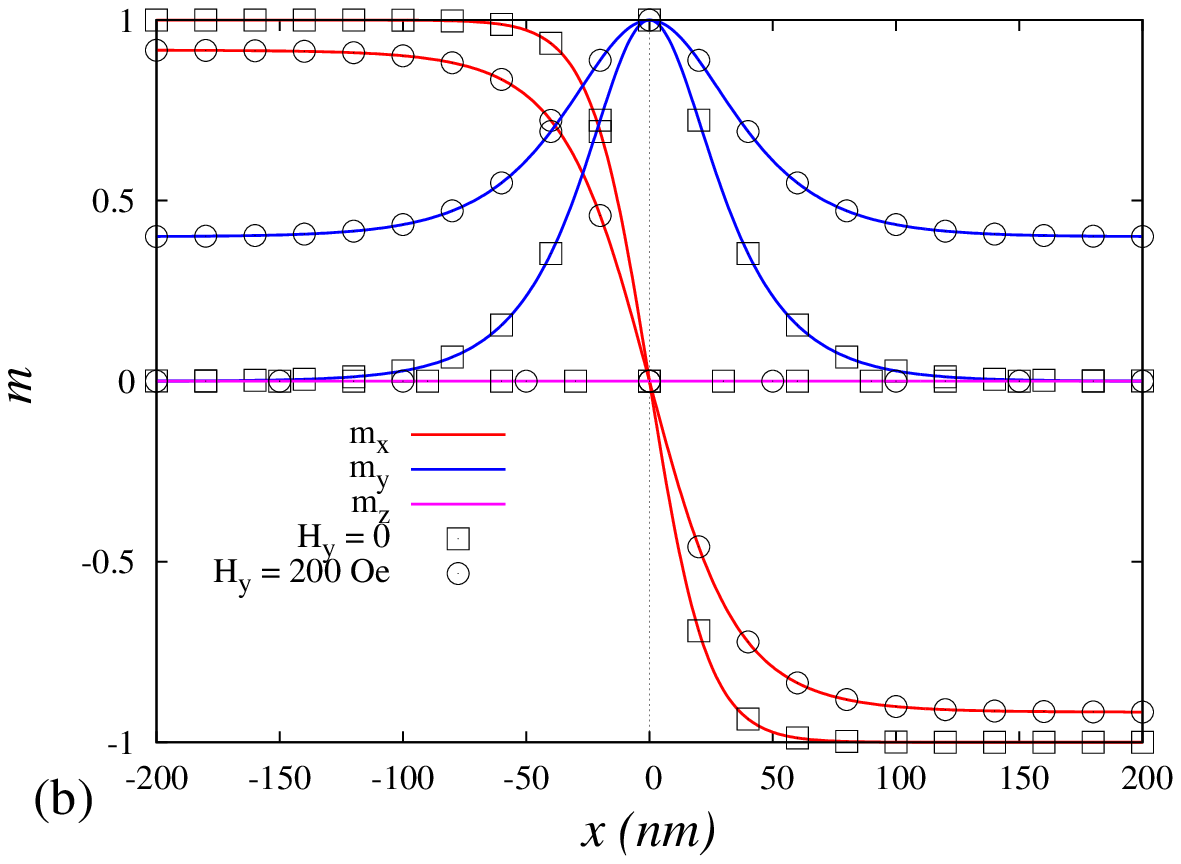}
\caption{(color online). (a) The spatial variation of $\theta$ for $H_y=0,200,400,500~Oe$ in the absence of driving field and current. (b) The spatial variation of the components of magnetization ($m_x,m_y$ and $m_z$) for $H_y=0,200~Oe$.}
\end{figure}

The static profile $\theta(x)$ and the normalised components of the magnetization ${\bf m(= M}/M_s)$ is given in Eqs.\eqref{statictheta} and \eqref{M1} respectively have been plotted in FIG. 3 for various values of transverse magnetic field.  When the strength of the transverse magnetic field is increased from 0 to 500 Oe, the value of $\theta$ in the left domain and right domain increases and decreases towards $\pi/2$  respectively as shown in the FIG.3(a).  The slope of $\theta$ at the center of the domain wall decreases while increasing the $H_y$ from 0 to 500 Oe. This implicitly refers to the expansion of the domain wall width. The spatial variation of the normalised components of the magnetization ${\bf m = M}/M_s$ is shown in FIG.3(b).  While the value of $M_z$ is zero everywhere, $M_x$ and $M_y$  are varied spatially. The broadening of the hump in the plot corresponding to $M_y$ due to the increase in $H_y$ from 0 Oe to 200 Oe and it confirms that, the increase in width of the domain wall.

\subsection{Trial function for the moving domain wall}

The static profiles for $\Phi$ and $\theta$ after applying transverse magnetic field, in the absence of current and driving field, are given by $\Phi(x)=0$ and $\theta(x)$(Eq.\eqref{statictheta}) respectively. Now we compute the trial functions of $\Phi$ and $\theta$ corresponding to the moving domain wall driven by the driving field and the current using the earlier static configuration. The applied current and driving field excert the spin-transfer torque and torque to the magnetic moments in the domain wall respectively and they create an energy difference between the two domains. In order to reduce the total energy of the nanostrip the domain wall moves. Though the presence of driving field can change the orientation angle $\theta$ in the left and right domains from the previous orientation angle $\theta_D$ and $(\pi-\theta_D)$ respectively and this variation can be neglected when the strength of the driving field is small. The another orientation angle $\Phi$ inside the two domains remains unchanged because the magnetic moments excite from the xy-plane and settle back to the same plane immediately due to damping. In the case of current along with transverse magnetic field, the orientation angels $\theta$ and $\Phi$ remains constant due to the absence of both the adiabatic and non-adiabatic spin-transfer torques within the domains. However the orientation angels $\theta$ and $\Phi$ varies inside the domain wall due to the torque and damping arise by the current and driving field. By neglecting the variation of $\theta$ and $\Phi$ inside the domains due to driving field and current, the trial functions for $\theta$ and $\Phi$ corresponding to the moving domain wall can be obtained by substituting the Eq.\eqref{statictheta} and $\Phi(x)=0$ in Eq.\eqref{trialtheta} and Eq.\eqref{trialphi} respectively and the results read

\begin{subequations}
\label{trialfunction1}
\begin{align}
\theta(x,t) &= 2\tan^{-1}\left[\frac{a_1+a_2 \exp\left( \frac{x-X(t)}{W(t)} \right)}{a_2+a_1 \exp\left( \frac{x-X(t)}{W(t)} \right)} \right],\label{theta1}\\
\Phi(x,t) &= \phi(t) ~U\left(\frac{x-X(t)}{W(t)} \right).\label{phi1}
\end{align}
\end{subequations}
The trial functions Eqs.\eqref{trialfunction1} represent the modelled solution for $\theta$ and $\Phi$ corresponding to the moving domain wall driven by current and driving field in the presence of transverse magnetic field. Using the above trial functions the unknown quantities such as excitation angle($\phi$), velocity($v$), width($W$) and displacement($X$) of the domain wall will be derived in the forthcoming section.

\subsection{The dynamical quantities of the domain wall}
On substituting $\frac{\partial\theta}{\partial t}$ from Eq.\eqref{llg2} in Eq.\eqref{llg1}, we obtain 
\begin{align}
\frac{\partial \Phi}{\partial t} = &\frac{1}{(1+\alpha^2)\sin\theta }\left\{~\frac{2A\gamma}{M_s}[\alpha\sin\theta~\frac{\partial^2 \Phi}{\partial x^2} + 2\alpha\cos\theta~\frac{\partial \theta}{\partial x}\frac{\partial \Phi}{\partial x}-\frac{\partial^2 \theta}{\partial x^2}+\sin\theta~\cos\theta~\left(\frac{\partial \Phi}{\partial x}\right)^2] \right.\nonumber\\
&\left.-2\alpha\gamma\pi M_s\sin2\Phi~\sin\theta + \frac{\gamma}{2}[H_k+4\pi M_s\sin^2\Phi]\sin2\theta+ \gamma H_d \sin\theta \right.\nonumber\\ 
&\left.- \alpha\gamma H_y\sin\Phi  - \gamma H_y\cos\Phi~\cos\theta+(1+\alpha\xi)b~\sin\theta ~\frac{\partial \Phi}{\partial x} + (\alpha-\xi)b~\frac{\partial \theta}{\partial x}\right\},
\label{eq1}
\end{align}
Similarly, by substituting $\frac{\partial\Phi}{\partial t}$ from Eq.\eqref{llg1} in \eqref{llg2}, we get
\begin{align}
\frac{\partial \theta}{\partial t} =&\frac{1}{(1+\alpha^2)}\left\{ ~\frac{2A\gamma}{M_s}[\sin\theta ~\frac{\partial^2 \Phi}{\partial x^2}+ 2\cos\theta~ \frac{\partial \theta}{\partial x}\frac{\partial \Phi}{\partial x} +\alpha\frac{\partial^2 \theta}{\partial x^2}-\alpha\sin\theta~\cos\theta~\left(\frac{\partial \Phi}{\partial x}\right)^2]\right.\nonumber\\ 
&\left.- 2\gamma\pi M_s\sin2\Phi~\sin\theta- \frac{\gamma\alpha}{2}\left[H_k+4\pi M_s\sin^2\Phi\right]\sin2\theta-\alpha\gamma H_d\sin\theta \right.\nonumber\\
& \left.- \gamma H_y\sin\Phi + \alpha\gamma H_y\cos\Phi~\cos\theta -(\alpha-\xi)b~\sin\theta~\frac{\partial \Phi}{\partial x} + (1+\alpha\xi)b~\frac{\partial \theta}{\partial x}\right\}. \label{eq2}
\end{align}
From Eqs.\eqref{theta1} and \eqref{phi1}, one can derive the following identities at $x = X(t)$.
\begin{subequations}
\label{identities}
\begin{align}
&\theta(X,t) = \pi\label{identity1},\\
&\frac{\partial \theta (X,t)}{\partial x} = \frac{1}{W(t)}~\sqrt{\frac{H_k-H_y}{H_k+H_y}}, \label{identity2} \\
&\frac{\partial^2 \theta (X,t)}{\partial x^2} = 0, \label{identity3}\\ 
&\frac{\partial^3 \theta (X,t)}{\partial x^3} = - \frac{H_k}{{W(t)^3}\left(H_k+{H_y}\right)} \sqrt{\frac{H_k-H_y}{H_k+H_y}}~,  \label{identity4} \\
&\frac{\partial \theta (X,t)}{\partial t} =-~\frac{1}{W(t)}\left(\frac{dX}{dt}\right)\sqrt{\frac{H_k-H_y}{H_k+H_y}}, \label{identity5}\\
&\Phi(X,t) = \phi(t),\label{identity6}\\
&\frac{\partial \Phi (X,t)}{\partial t} = \frac{d\phi(t)}{dt},\label{identity7}\\
&\frac{\partial \Phi (X,t)}{\partial x} = \frac{\partial^2 \Phi (X,t)}{\partial x^2} =  \frac{\partial^2 \Phi (X,t)}{\partial x \partial t} = 0.\label{identity8} 
\end{align}
\end{subequations}
After reducing Eq.\eqref{eq1} at $x=X(t)$ and substituting Eqs.\eqref{identity1}-\eqref{identity8} in it, we get
\begin{align}
(1+\alpha^2)\frac{d\phi(t)}{dt} = \gamma (H_d-\alpha[2\pi M_s\sin2\phi + H_y \sin\phi]) 
 +\frac{(\alpha-\xi)b}{W(t)} \sqrt{\frac{H_k-H_y}{H_k+H_y}}. \label{dphi}
\end{align}
Similarly, the velocity of the domain wall is obtained from Eq.\eqref{eq2} as
\begin{align}
v(t)=\frac{dX}{dt} = \frac{\gamma W(t) }{(1+\alpha^2)}\sqrt{\frac{H_k+H_y}{H_k-H_y}} {\left[2\pi M_s\sin2\phi + \alpha H_d + H_y \sin\phi \right]} - b\left(\frac{1+\alpha\xi}{1+\alpha^2}\right). \label{v}
\end{align} 
The width of the domain wall is obtained after differentiating Eq.\eqref{eq1} with respect to $x$ and reducing the differentiated equation at $x=X(t)$ using Eqs.\eqref{identity1}-\eqref{identity8}.
\begin{equation}
W(t) ={W_0}{\sqrt{\frac{H_k}{H_y+H_k}}} \left[1 + \frac{4\pi M_s}{H_k} \sin^2\phi -\frac{H_y}{H_k} \cos\phi \right]^{-\frac{1}{2}}, \label{W}
\end{equation}
and the ratio of the width to the initial width namely the width ratio of the domain wall is given by,
\begin{align}
\frac{W(t)}{W(0)}=\sqrt{1-\frac{H_y}{H_k}} \left[1 + \frac{4\pi M_s}{H_k} \sin^2\phi -\frac{H_y}{H_k} \cos\phi \right]^{-\frac{1}{2}}.\label{Wratio}
\end{align}
Eqs. \eqref{dphi}, \eqref{v} and \eqref{W} represent the excitation angle($\phi$), velocity($v$) and width($W$) of the domain wall driven by driving field and current in the presence of a transverse magnetic field.  From the equations \eqref{v} and \eqref{W}, one can understand that the velocity and the width of the domain wall are the function of excitation angle. Hence, we need to solve the Eq.\eqref{dphi} in order to understand the domain wall motion. The Eq.\eqref{dphi} is a highly nontrivial nonlinear evolution equation and solving it analytically in the present form is extremely difficult.  But it is possible to find out the exact analytical solution of Eq.\eqref{dphi} only when $H_d\neq 0,b=H_y=0$ and $H_y\neq 0,b=H_d=0$ and for the other cases, the Eq.\eqref{dphi} can be solved using small angle approximation technique which would be discussed in the forthcoming sections.

\subsection{Exact analytical solution for the domain wall motion}
\begin{figure}[!hbtp]
\centering\includegraphics[angle=0,width=0.33\linewidth]{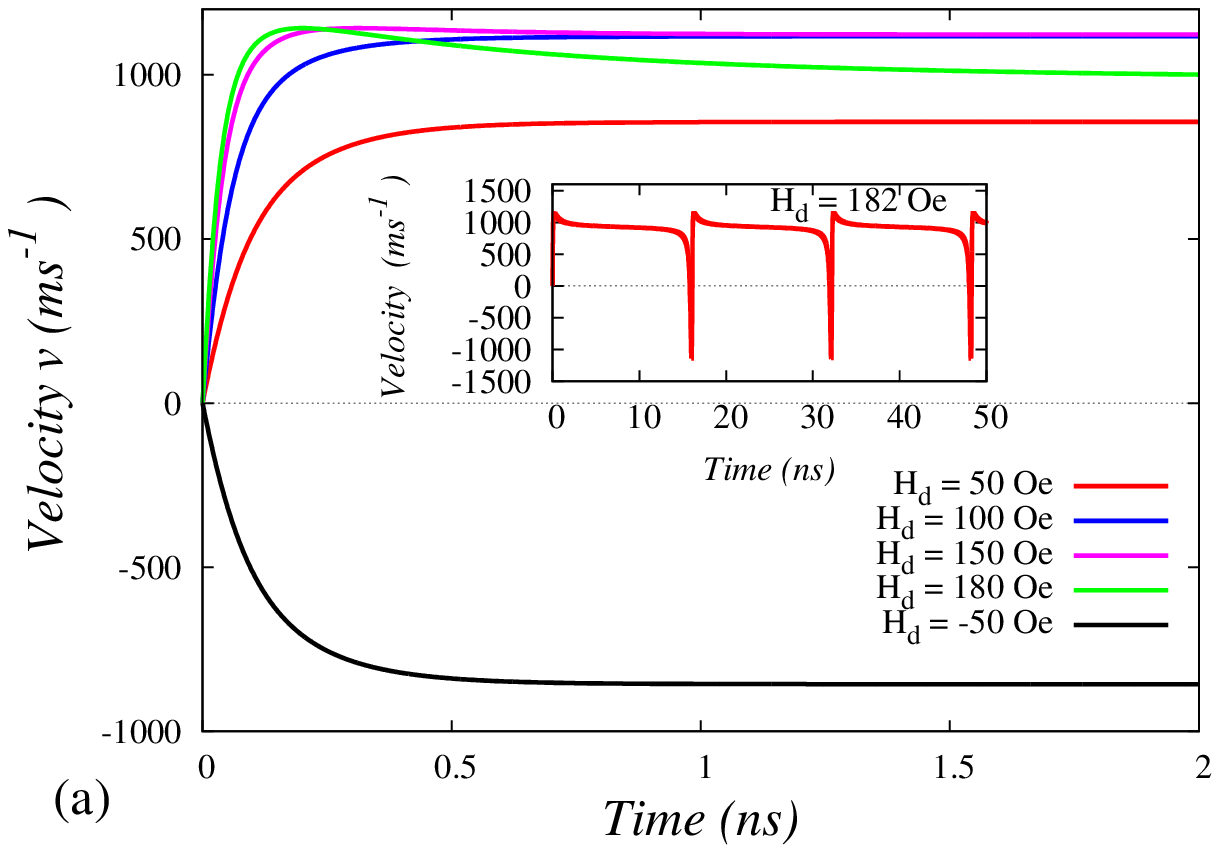}~\includegraphics[angle=0,width=0.33\linewidth]{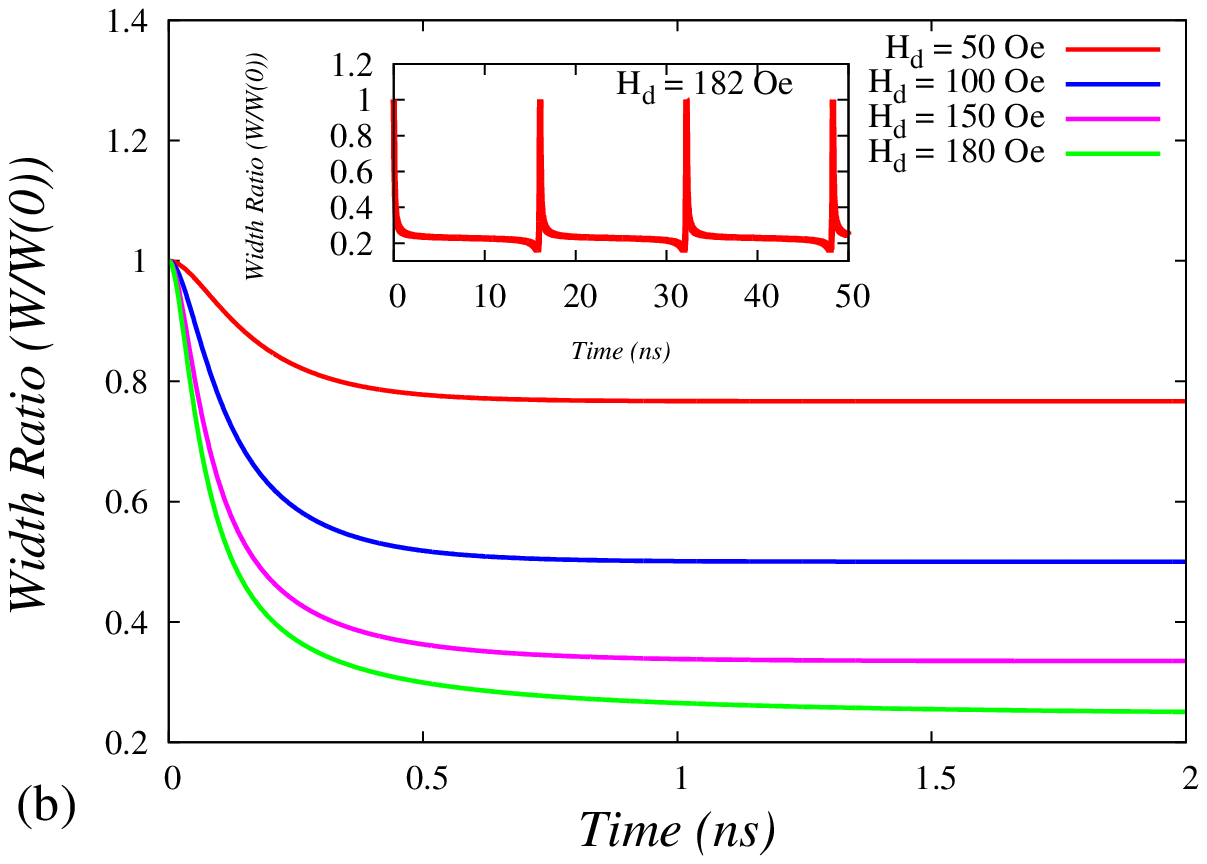}~\centering\includegraphics[angle=0,width=0.33\linewidth]{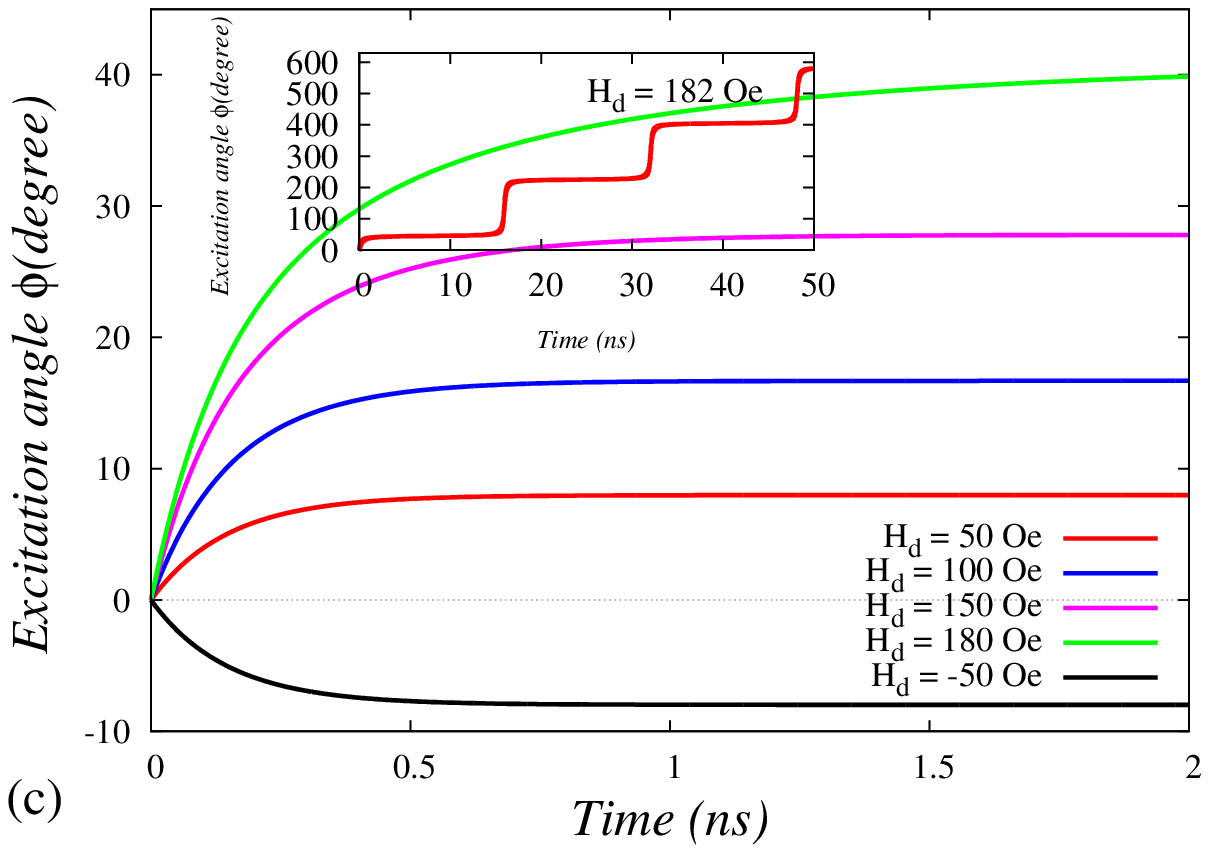}
\caption{(color online). The variation of (a) velocity, (b) width ratio and (c) excitation angle of the domain wall with respect to time driven by different driving fields below Walker limit(181.6 Oe).  The corresponding inset figures show the variation above Walker limit.}
\end{figure}
\subsubsection{In the presence of driving field alone($H_d\neq 0$ and $b=H_y=0$):}
To study the dynamics of the domain wall in the presence of driving field and absence of current and transverse magnetic field, Eq.\eqref{dphi} is rewritten as
\begin{align}
(1+\alpha^2)\frac{d\phi(t)}{dt} = \gamma (H_d-2\pi\alpha M_s\sin2\phi ).\label{dphiHd}
\end{align}

The solution $\phi(t)$ can be obtained by integrating the Eq.\eqref{dphiHd}, is given as

\begin{align}
\int_{\phi(0)}^{\phi(t)}\frac{d\phi}{H_d-2\pi\alpha M_s\sin2\phi } = \int_0^t\frac{\gamma}{1+\alpha^2} dt. \label{intHd}
\end{align}
For the initial condition is $\phi(0)=0$, the excitation angle $\phi(t)$ is solved from Eq.\eqref{intHd} as
\begin{align}
\phi(t) &= \tan^{-1}\left\{\frac{(G+2\pi\alpha M_s)\exp\left\{-\left(\ln\left[\frac{2\pi\alpha M_s+G}{2\pi\alpha M_s-G}\right]+\frac{2\gamma G }{1+\alpha^2}t\right)\right\}+G-2\pi\alpha M_s}{H_d\left(\exp\left\{-\left(\ln\left[\frac{2\pi\alpha M_s+G}{2\pi\alpha M_s-G}\right]+\frac{2\gamma G }{1+\alpha^2}t\right)\right\}-1\right)} \right\},\nonumber\\ 
&~~~~~~~~~~~\rm{when}~H_d^2<4\pi^2\alpha^2 M_s^2, \label{phiHd}\\
\phi(t) &= \tan^{-1}\left\{\frac{2\pi\alpha M_s}{H_d}+\sqrt{1-\left(\frac{2\pi\alpha M_s}{H_d}\right)^2}\tan\left(\tan^{-1}\left[\frac{-2\pi\alpha M_s}{\sqrt{H_d^2-4\pi^2\alpha^2 M_s^2}}\right]+\frac{\gamma\sqrt{H_d^2-4\pi\alpha^2 M_s^2}}{1+\alpha^2}t\right) \right\},\nonumber\\ 
&~~~~~~~~~~~\rm{when}~H_d^2>4\pi^2\alpha^2 M_s^2, \label{phiHd_above_Walker}
\end{align}
where, $G=\sqrt{4\pi^2\alpha^2 M_s^2-H_d^2}$. 
The presence of tangent function in Eq.\eqref{phiHd_above_Walker} implies the periodic variation in $\phi$ when the magnitude of driving field is above the value of $2\pi\alpha M_s$(=181.6 Oe) and it is referred to the Walker limit of the field\cite{Schryer}.  The velocity(Eq.\eqref{v}) and the width(Eq.\eqref{W}) of the domain wall can be obtained for the driving fields corresponding to below and above the Walker limit of field from Eqs.\eqref{phiHd} and \eqref{phiHd_above_Walker} respectively.  From Eq.\eqref{phiHd}, we can show that when $t\to\infty$, the $\phi(t)$ reaches the constant value, which can be called as saturated excitation angle($\phi_s$) and it can derived as
\begin{align}
\phi_s=\phi(\infty) = \tan^{-1}\left(\frac{2\pi\alpha M_s}{H_d}-\sqrt{\left(\frac{2\pi\alpha M_s}{H_d}\right)^2-1}\right)=\mathrm{constant}. \label{saturatedphiHd}
\end{align}
Correspondingly, the saturated velocity($v_s=v(\infty)$) and the saturated width($W_s=W(\infty)$) can be calculated by substituting Eq.\eqref{saturatedphiHd} in Eqs.\eqref{v} and \eqref{W} respectively.  From Eq.\eqref{saturatedphiHd} we can observe that at $t=\infty$, $d\phi/dt=0$. 

The velocity of the domain wall can also be obtained by reducing the Eq.\eqref{llg2} at $x=X$  and using Eqs.\eqref{identities} as
\begin{align}
v(t)  = \frac{dX}{dt} = \frac{W(t)}{\alpha}\sqrt{\frac{H_k+H_y}{H_k-H_y}} \left[\gamma H_d-\frac{d\phi}{dt}\right] - b~\frac{\xi}{\alpha}.\label{satvelocity}
\end{align}
By substituting $d\phi/dt=0$ for $t=\infty$ in Eq.\eqref{satvelocity}, we get the saturated velocity of the domain wall in the absence of current and transverse magnetic field
\begin{align}
v_s=v(\infty) = \frac{\gamma H_d W_s}{\alpha}.\label{satvelocity1}
\end{align} 
For the understanding of domain wall motion, the quantities $v(t)$, $W(t)/W(0)$ and $\phi(t)$ are plotted in the figures 4(a-c) for the driving fields below the Walker limit(181.6 Oe) and the corresponding inset figures exhibit the oscillatory behaviour of the domain wall for the driving field above Walker limit. 

From FIG.4(a), it is observed that while applying the driving field along the positive x-direction, the domain wall starts to move slowly in the same direction.  Also, the initial velocity $v(0)$ is consistent with the expression $v(0)=\alpha\gamma W_0 H_d/(1+\alpha^2)\approx 0$ derived from Eq.\eqref{v}. After a time around t=0.5 ns, the velocity of the domain wall gets saturated when the reduction in Zeeman energy of the magnetic strip equal to the dissipation of energy through damping\cite{Zhang1}, and the corresponding saturated value varies depending on the strength of the driving field which is also consistent with the equation \eqref{satvelocity1}.  The plots corresponding to $H_d$=50 Oe and -50 Oe imply that the direction of the velocity of the wall is same as the direction of the driving field and the magnitude of the velocity is independent of the direction of the driving field.
The saturated velocity increases with the increase of driving field when its strength is lower and decreases with the increase of driving field when the strength of driving field reaches the Walker limit is shown in FIG.4(a). This can be understood from the expression of the saturated velocity which is proportional to the product of driving field, and saturated width and the saturated width decreases with the increase of driving field as shown in the FIG.4(b). Initially, the width ratio assumes to be unity and decreases with respect to time and reaches its saturation $W_s/W(0)$ at around t=0.5 ns for different driving fields.  The time variation of the width ratio is irrespective of the direction of the driving field.   The saturated width ratio $W_s/W(0)$ decreases with increase of driving field implies that the saturated width $W_s$ decreases when the driving field increases as shown in FIG.4(b). The excitation angle $\phi$ starts from zero and reaches its saturation $\phi_s$ around time t=0.5 ns for different driving field is shown in FIG.4(c). The variation in the sign of the excitation angle represents that the the magnetic moments in the domain wall excite upwards or downwards with respect to the plane of the nanostrip corresponding to positive or negative direction of the driving field respectively. The decrease in the width ratio(see FIG.4(b)) and increase in the excitation angle((see FIG.4(c)) with respect to time implicitly indicate the increase in distortion observed in the domain wall for all driving fields below the Walker limit. The distortion of the domain wall is associated with damping and it increases with time upto around 0.5 ns after that the rate of decrease in zeeman energy balances the rate of damping in energy dissipation and the wall starts to move with constant velocity.  

\subsubsection{In the presence of transverse magnetic field alone($H_y\neq 0$ and $b=H_d=0$):}

The variation of the excitation angle of the domain wall after applying the transverse magnetic field is obtained by substituting $b=H_d=0$ in Eq.\eqref{dphi} is given as

\begin{align}
(1+\alpha^2)\frac{d\phi(t)}{dt} = -\alpha\gamma (2\pi M_s\sin2\phi + H_y \sin\phi). \label{dphiHy}
\end{align}
Eq.\eqref{dphiHy} can be integrated as
\begin{align}
\int_{\phi(0)}^{\phi(t)} \frac{d\phi}{2\pi M_s\sin2\phi + H_y \sin\phi} = -\int_0^t\frac{\alpha\gamma}{1+\alpha^2} dt.\label{intHy}
\end{align}
We obtain the following equation after integrating the Eq.\eqref{intHy}.
\begin{align}
\left(\tan\frac{\phi}{2}\right)^{H_y}\left(\frac{H_y+4\pi M_s\cos\phi}{\sin\phi}\right)^{4\pi M_s}=&\exp\left(4\pi M_s\ln\left[\frac{H_y+4\pi M_s\cos\phi(0)}{\sin\phi(0)}\right]\right)\nonumber\\
&\exp\left(H_y\ln\left[\tan\frac{\phi(0)}{2}\right]\right)\exp\left(-\frac{\alpha\gamma(H_y^2-16\pi^2 M_s^2)}{1+\alpha^2}t\right).\label{intHy1}
\end{align}
By applying the initial condition $\phi(0)=0$ in the above equation (Eq.\eqref{intHy1}), $\phi(t)$ is obtained as

\begin{align}
\phi(t) = 0. \label{phiHy}
\end{align}
On substituting Eq.\eqref{phiHy} in Eqs.\eqref{v} and \eqref{W}, we get
\begin{align}
v(t) = 0~\mathrm{and}~W(t)=W_0/\sqrt{1-\left(\frac{H_y}{H_k}\right)^2}=\mathrm{constant}.  \label{vW_Hyzero}
\end{align}
Eqs.\eqref{phiHy} and \eqref{vW_Hyzero} represent that there is no variation in the excitation angle, velocity and width of the domain wall with respect to time only when a transverse magnetic field is applied. 

\subsection{Analytical solution for the domain wall motion using small angle approximation}

The analytical solution for the excitation angle $\phi$ under the small angle approximation are valid only for below the Walker limit. However for above the Walker limit, the domain wall would exhibits oscillatory behaviour which has been observed from Eq.\eqref{phiHd_above_Walker} in the previous section. In the forthcominng section, we discuss the analytical solution for the excitation angle, the velocity and the width of the domain wall in presence of transverse magnetic field along with driving field and current.

\subsubsection{The general case($b,H_d,H_y\neq 0$)}
To understand the motion of the domain wall driven by the driving field and current in the presence of transverse magnetic field, the excitation angle of the wall is obtained by rewriting the Eq.\eqref{dphi} using the small angle approximations $\sin\phi \approx \phi$ and $\cos\phi\approx 1$ as
\begin{align}
(1+\alpha^2)\frac{d\phi(t)}{dt} = \gamma H_d - \alpha\gamma(4\pi M_s+H_y)\phi+\frac{(\alpha-\xi)b}{W_0}\left[{\left(1-\frac{H_y}{H_k}\right)\left(1+\frac{4\pi M_s\phi^2-H_y}{H_k}\right)}\right]^{1/2}.\label{dphi1}
\end{align}
Since, $\phi$ is very small, Eq.\eqref{dphi1} is rewritten after making the approximation $[{1+{(4\pi M_s\phi^2-H_y)}/{H_k}}]^{1/2}\approx [{1+{(4\pi M_s\phi^2-H_y)}/{2H_k}}]$ as
\begin{align}
\frac{d\phi(t)}{dt} = E~\phi(t)^2 - F~\phi(t) + G, \label{dphi2}
\end{align}
where E, F and G are given by
\begin{subequations}
\begin{align}
&E = \frac{2\pi M_s(\alpha-\xi)b\sqrt{1-\frac{H_y}{H_k}}}{H_k W_0(1+\alpha^2)},\label{E}\\
&F = \frac{\alpha\gamma(4\pi M_s+H_y)}{1+\alpha^2},\label{F}\\
&G = \frac{1}{1+\alpha^2}\left[\gamma H_d+\frac{(\alpha-\xi)b}{W_0}\sqrt{1-\frac{H_y}{H_k}}\left(1-\frac{H_y}{2H_k} \right) \right].\label{G}
\end{align}
\end{subequations}
Eq.\eqref{dphi2} is in the form of the well known Riccati equation, the solution can be written as
\begin{align}
\phi(t) = \frac{1}{Z(t)}+\phi_p, \label{saa1}
\end{align}
where $\phi_p$ is a particular solution of Eq.\eqref{dphi2} and $Z(t)$ is yet to be determined.  Since the time independent solution of Eq.\eqref{dphi2} is a particular solution of the same equation, $\phi_p$ is derived after substituting $\phi$=$\phi_p$=constant in Eq.\eqref{dphi2}.  The result reads
\begin{align}
&\phi_p =\phi_{\pm}= \frac{\alpha\gamma H_k(4\pi M_s+H_y)W_0}{4\pi M_s(\alpha-\xi)b\sqrt{1-\frac{H_y}{H_k}}}\nonumber\\
&~~~~~~~~\left\{1\pm\left[1-\frac{8\pi M_s(\alpha-\xi)b\sqrt{1-\frac{H_y}{H_k}}}{\alpha^2\gamma^2H_k(4\pi M_s+H_y)^2 W_0}\left(\gamma H_d+\frac{(\alpha-\xi)b}{W_0}\sqrt{1-\frac{H_y}{H_k}}\left(1-\frac{H_y}{2H_k} \right) \right) \right]^{\frac{1}{2}}  \right\}. \label{phi_pm}
\end{align}
Where $\phi_+$ and $\phi_-$ are two time independent solutions of Eq.\eqref{dphi2}. On substituting Eq.\eqref{saa1} in Eq.\eqref{dphi2}, $Z(t)$ is derived as
\begin{align}
\frac{dZ(t)}{dt} = (F-2E \phi_p)Z(t) - E. \label{saa2}
\end{align}
Integrating Eq.\eqref{saa2}, we get
\begin{align}
Z(t) = \frac{E}{F-2E\phi_p} + C_3 \exp\left[(F-2E\phi_p)t \right], \label{saa3}
\end{align}
where $C_3$ is the constant of integration.  After substituting Eq.\eqref{saa3} in Eq.\eqref{saa1}, we obtain
\begin{align}
\phi(t) = \left[\frac{E}{F-2E\phi_p}+C_3 \exp\left[(F-2E\phi_p)t \right] \right]^{-1} + \phi_p .\label{saa4}
\end{align}
The constant of integration $C_3$ is evaluated using the initial condition $t=0,\phi=\phi_0$ in Eq.\eqref{saa4}.  The result reads
\begin{align}
C_3 = -\left(\frac{E}{F-2E\phi_p}-\frac{1}{\phi_0-\phi_p} \right). \label{C2}
\end{align}
By substituting the equations \eqref{C2}, $F=E(\phi_++\phi_-)$ and $\phi_p=\phi_+$ in Eq.\eqref{saa4}, we get
\begin{align}
\phi(t) = \left\{\phi_+ + \frac{(\phi_0-\phi_+)(\phi_- -\phi_+)}{(\phi_0-\phi_+)-(\phi_0-\phi_-)\exp(E(\phi_- -\phi_+)t)} \right\}, ~\phi_-<\phi_+,~b\neq 0.\label{phi}
\end{align}
Eq.\eqref{phi} gives the analytical solution for the excitation angle of the domain wall under small angle approximation in the presence of a transverse magnetic field($H_y$), driving field($H_d$) and spin-transfer torque($b$).  
The velocity and width of the domain wall can be determined from Eq.\eqref{phi} using Eqs.\eqref{v} and \eqref{W} respectively.  Since, $E(\phi_- -\phi_+)<0$, one can show from Eq.\eqref{phi} that the excitation angle $\phi$ reaches the saturated value $\phi_s$  when $t\to \infty$.
\begin{align}
\phi(\infty)=\phi_s=\phi_-=\mathrm{constant}.\label{phi_s}
\end{align}
Eq.\eqref{phi_s} implies that the excitation angle saturates as time passes and simillarly the velocity(Eq.\eqref{v}) and the width(Eq.\eqref{W}) of the domain wall reach the saturated values $v_s$ and $W_s$ respectively because they are explicitly depend on $\phi$.  By substituting $d\phi/dt$=0 and $W(t)=W_s$ in Eq.\eqref{satvelocity}, we can derive the saturated velocity as
\begin{align}
v_s  = \frac{\gamma H_d W_s}{\alpha}\sqrt{\frac{H_k+H_y}{H_k-H_y}}  - b~\frac{\xi}{\alpha}.\label{satvelocity2}
\end{align}
From the saturated velocity expression, we observed that the saturated velocity of a domain wall is high  only when the current and the driving field are applied in the opposite directions whereas it is low for the current and the driving field are applied in the same directions. In the absence of driving field, the transverse magnetic field does not influence the saturated velocity of the domain wall and it is controlled by the adiabatic spin-transfer torque $b$.  In the presence of driving field, the saturated velocity increases with the transverse magnetic field.  It implies the saturated velocity of a domain wall cannot be increased by the transverse magnetic field in the absence of driving field.

\subsubsection{For the case when $H_d,H_y\neq 0$ and $b=0$}

When the current is switched off($b=0$), the excitation angle $\phi$ given in Eq.\eqref{phi} is undetermined. To find the solution of the excitation angle in the presence of transverse magnetic field, driving field and absence of current, Eq.\eqref{dphi2} is written after substituting $b=0$ as
\begin{align}
\frac{d\phi}{dt} =  \frac{\gamma H_d}{(1+\alpha^2)}- \frac{\alpha\gamma(4\pi M_s+H_y)}{(1+\alpha^2)}~\phi(t) , \label{eq1_zerocurrent}
\end{align}
On integrating Eq.\eqref{eq1_zerocurrent}, we get
\begin{align}
\ln\left( { H_d}- {\alpha(4\pi M_s+H_y)\phi}\right) = - \frac{\alpha\gamma(4\pi M_s+H_y)t}{(1+\alpha^2)}  + C_4, \label{eq2_zerocurrent}
\end{align}
where, the constant of integration $C_4$ is obtained using the initial condition $t=0,\phi=\phi_0$.
\begin{align}
C_4 = \ln\left( {\gamma H_d}- {\alpha(4\pi M_s+H_y)\phi_0} \right).\label{C3}
\end{align}
After substituting $C_4$ in Eq.\eqref{eq2_zerocurrent}, the value of $\phi$ for the domain wall driven by driving field in the presence of transverse magnetic field is derived as follows
\begin{align}
\phi(t) = \frac{H_d}{\alpha(4\pi M_s+H_y)}+\left(\phi_0- \frac{H_d}{\alpha(4\pi M_s+H_y)}\right)\exp\left(- \frac{\alpha\gamma(4\pi M_s+H_y)t}{1+\alpha^2} \right). \label{phi_zerocurrent}
\end{align}
The corresponding velocity and width of the domain wall can be found from Eqs.\eqref{v} and \eqref{W} respectively using Eq.\eqref{phi_zerocurrent}.  The saturated excitation angle($\phi_s$) is given from Eq.\eqref{phi_zerocurrent} at $t=\infty$ as
\begin{align}
\left[\phi_s\right]_{b=0} = \frac{H_d}{\alpha(4\pi M_s+H_y)}. \label{phis_zerocurrent}
\end{align}
The Eq.\eqref{phis_zerocurrent} shows that the increase in transverse magnetic field decreases the magnitude of saturated excitation angle. This implies that the excitation of the domain wall from the plane of strip can be controlled by the transverse magnetic field. 
\subsection{Displacement of the domain wall}

The displacement of the domain wall $X(t)$ is obtained by integrating Eq.\eqref{satvelocity} with respect to time and using the initial condition $\phi(0)=0,X(0)=0$.
\begin{align}
X(t) =\int_0^t v(t)dt= \frac{1}{\alpha}\sqrt{\frac{H_k+H_y}{{H_k}-H_y}}\int_0^t W(t)\left(\gamma H_d-\frac{d\phi}{dt} \right)dt - \int_0^t b\frac{\xi}{\alpha} dt. \label{disp1}
\end{align}
Eq.\eqref{disp1} can be rewritten using Eq.\eqref{W} as
\begin{align}
X(t) = \frac{W_0}{\alpha\sqrt{1-\frac{H_y}{H_k}}}\int_0^t \frac{\left(\gamma H_d-\frac{d\phi}{dt} \right)dt}{\sqrt{1 + \frac{4\pi M_s}{H_k} \sin^2\phi -\frac{H_y}{H_k}\cos\phi }} - b\frac{\xi}{\alpha}\int_0^t  dt. \label{disp2}
\end{align}
As it is difficult to integrate the Eq.\eqref{disp2} in its present form, we implement the small angle approximations $\sin\phi \approx \phi$ and $\cos\phi\approx 1$ and the result reads
\begin{align}
X(t) = \frac{W_0}{\alpha\sqrt{1-\frac{H_y}{H_k}}}\left(\gamma H_d I_1-I_2\right) - b\frac{\xi}{\alpha}\int_0^t  dt. \label{disp3}
\end{align}
Where, the integrals $I_1$ and $I_2$ are given by,
\begin{align}
I_1 = \int_0^t \frac{dt}{\sqrt{1+\frac{4\pi M_s}{H_k}\phi^2-\frac{H_y}{H_k}}},~~~I_2 = \int_0^\phi \frac{d\phi}{\sqrt{1+\frac{4\pi M_s}{H_k}\phi^2-\frac{H_y}{H_k}}}.\label{I}
\end{align}
The integrals $I_1$ and $I_2$ are evaluated using Mathematica 8.0 and the results read
\begin{subequations}
\label{Int}
\begin{align}
I_1 = \frac{\sqrt{H_k}}{EF_1F_2(\phi_+-\phi_-)}&\bigg\{EF_1(\phi_+-\phi_-)t\nonumber\\
&+F_1\log\left[F_3^{-1}((\phi_--\phi_+F_3)(H_k-H_y)-4\pi M_s(F_3-1)\phi_-^2\phi_++F_2F_4) \right]\nonumber\\
&+F_1\log\left[F_3^{-1}((\phi_--\phi_+F_3)(H_y-H_k)+4\pi M_s(F_3-1)\phi_+^2\phi_-+F_1F_4) \right] \bigg\},\label{Int1}\\
I_2 = \sqrt{\frac{H_k}{4\pi M_s}}&\log\left[\left(1-\frac{H_y}{H_k}\right)^{-{1/2}}\left(\sqrt{\frac{4\pi M_s}{H_k}}\phi+\sqrt{1+\frac{4\pi M_s}{H_k}\phi^2-\frac{H_y}{H_k}}\right)\right],\label{Int2}
\end{align}
\end{subequations}
where,
\begin{subequations}
\label{F}
\begin{align}
&F_1 = \sqrt{H_k-H_y+4\pi M_s \phi_+^2},\label{F1}\\
&F_2 = \sqrt{H_k-H_y+4\pi M_s \phi_-^2},\label{F2}\\
&F_3 = \exp\left\{E(\phi_+-\phi_-)t\right\},\label{F3}\\
&F_4 = \sqrt{(H_k-H_y)(\phi_--\phi_+F_3)^2+4\pi M_s(F_3-1)^2\phi_+^2\phi_-^2} \label{F4}
\end{align}
\end{subequations}
and $E$ is given by \eqref{E}. By substituting the Eqs.\eqref{Int} in Eq.\eqref{disp3} gives the expression for the displacement of the domain wall. The above result is valid only for b$\neq$0, if b=0, the integrals $I_1$ and $I_2$ are undetermined. Hence, to calculate the displacement of the domain wall for the case of b=0, the integrals $I_1$ and $I_2$ in Eq.\eqref{disp3} are evaluated using the same initial condition $\phi(0)=0,X(0)=0$  with the excitation angle given by Eq.\eqref{phi_zerocurrent} instead of Eq.\eqref{phi}. Therefore, the integral $I_1$ is given by
\begin{align}
I_1 = &\frac{\sqrt{H_k}}{G_2\gamma}\bigg\{\alpha\gamma(4\pi M_s+H_y)t+(1+\alpha^2)\nonumber\\
&\log\left[G_1^{-1}(4\pi M_s(G_1-1)H_d^2+\alpha G_1(H_y+4\pi M_s)[\alpha(H_y-H_k)(H_y+4\pi M_s)-\sqrt{H_k}G_2G_3] ) \right] \bigg\},\label{Int3}
\end{align}
and the integral $I_2$ is same as given in Eq.\eqref{Int2} except that $\phi$ is obtained from Eq.\eqref{phi_zerocurrent}.
 where
\begin{align}
&G_1 = \exp\left\{\frac{\alpha\gamma(H_y+4\pi M_s)t}{1+\alpha^2} \right\},\nonumber\\
&G_2 = \sqrt{4\pi M_s H_d^2+(H_k-H_y)(4\pi M_s+H_y)^2\alpha^2},\nonumber\\
&G_3 = \frac{\sqrt{4(F_1-1)^2\pi M_s H_d^2-F_1^2(H_y-H_k)(H_y+4\pi M_s)^2\alpha^2}}{\alpha F_1\sqrt{H_k}(H_y+4\pi M_s)}.
\end{align}


\section{Confirmation of analytical solutions with numerical results and discussions}

In this section, the excitation angle, velocity, width ratio and displacement of the domain wall are obtained from the approximated analytical solutions Eqs.\eqref{phi} and \eqref{phi_zerocurrent} are verified with the corresponding numerical results obtained by integrating the dynamical equation \eqref{dphi} using Runge-Kutta-4 algorithm with the initial condition $\phi(0)=0$ and by using the experimentally measured values of the material parameters of Cobalt nanostripes as given by $M_s=14.46\times10^5~Am^{-1},~4\pi M_s=1.8\times10^4~Oe,~A=2\times10^{-11}~Jm^{-1},~\gamma=1.9\times10^7~Oe^{-1}s^{-1},~H_k = 500~Oe$,~$P=0.35$ and $\xi=0.01$\cite{Li,Zhang}.  
\subsection*{Case A: Effect of the current in the Domain Wall motion($b,c\neq 0;~H_d=H_y=0$)}
\begin{figure}[!hbtp]
\centering\includegraphics[angle=0,width=0.5\linewidth]{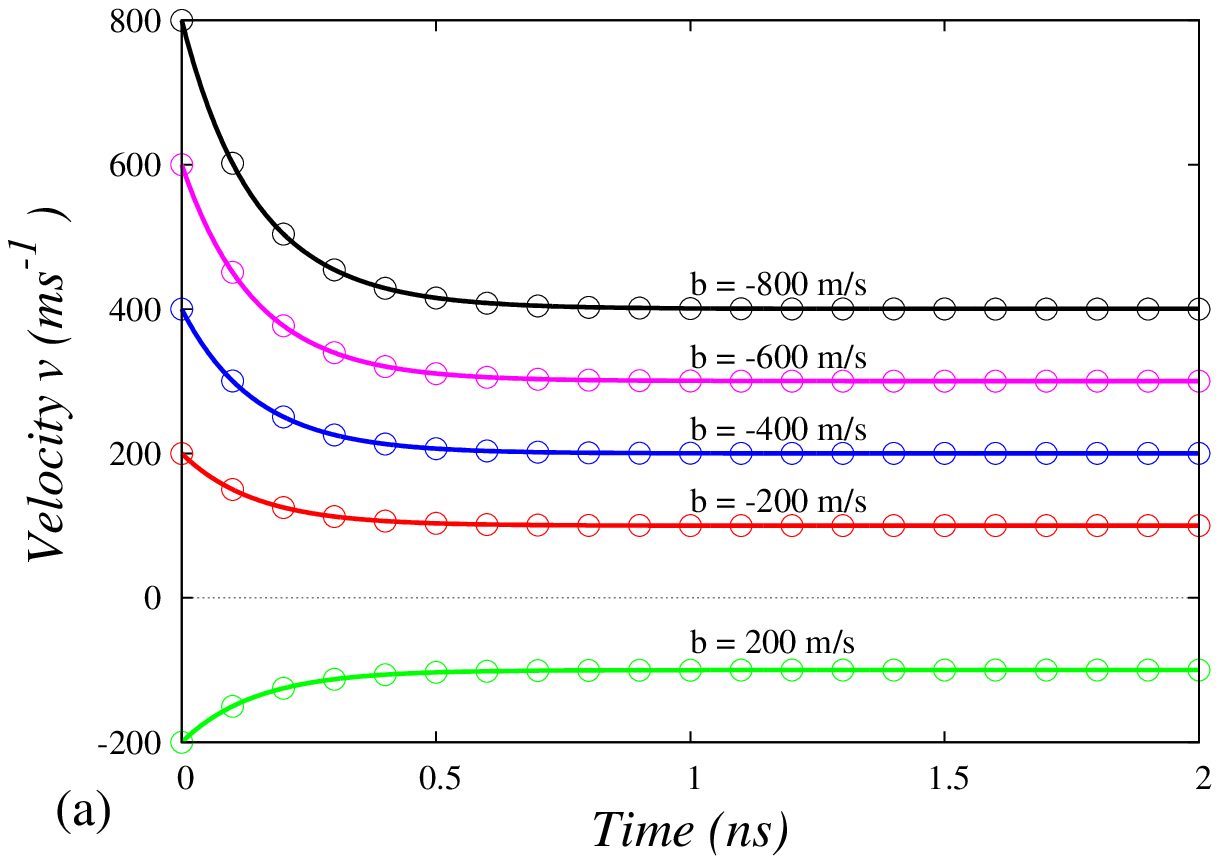}~\includegraphics[angle=0,width=0.5\linewidth]{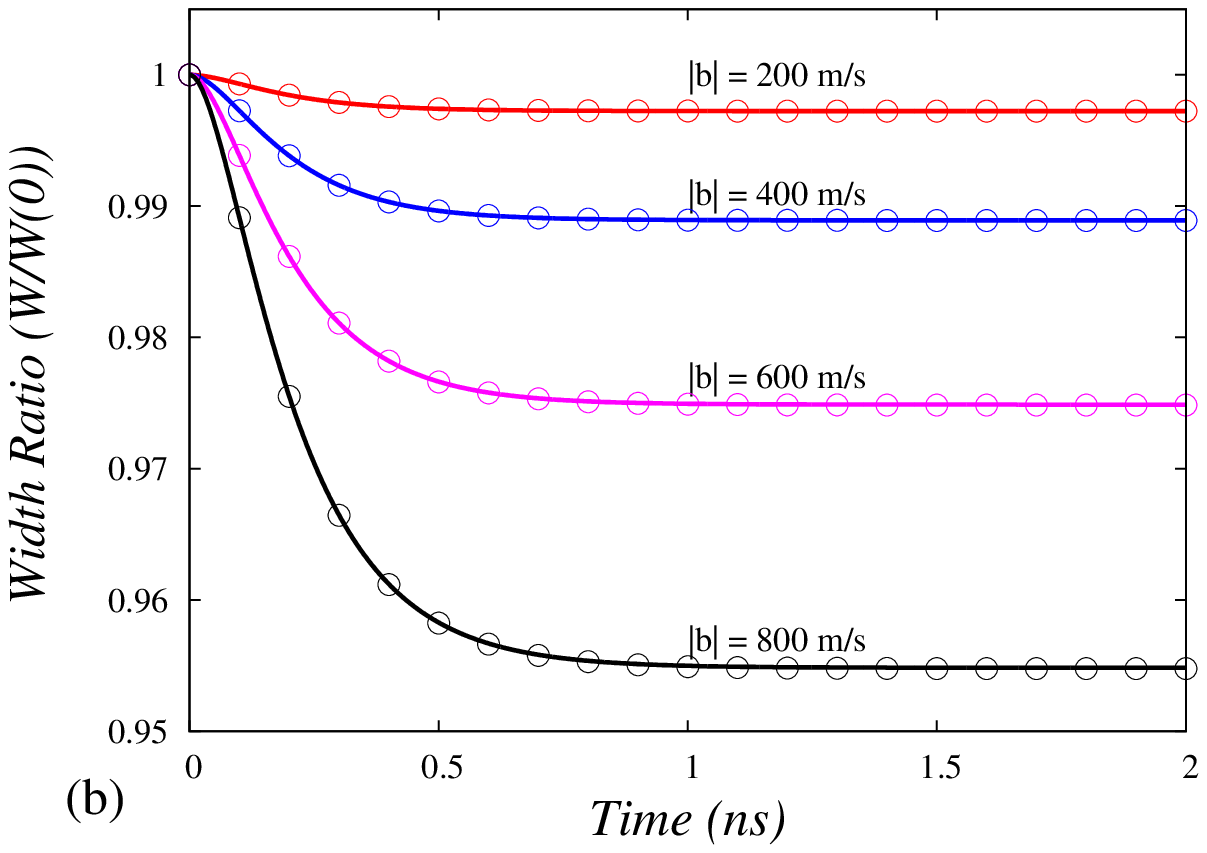}\\
\centering\includegraphics[angle=0,width=0.5\linewidth]{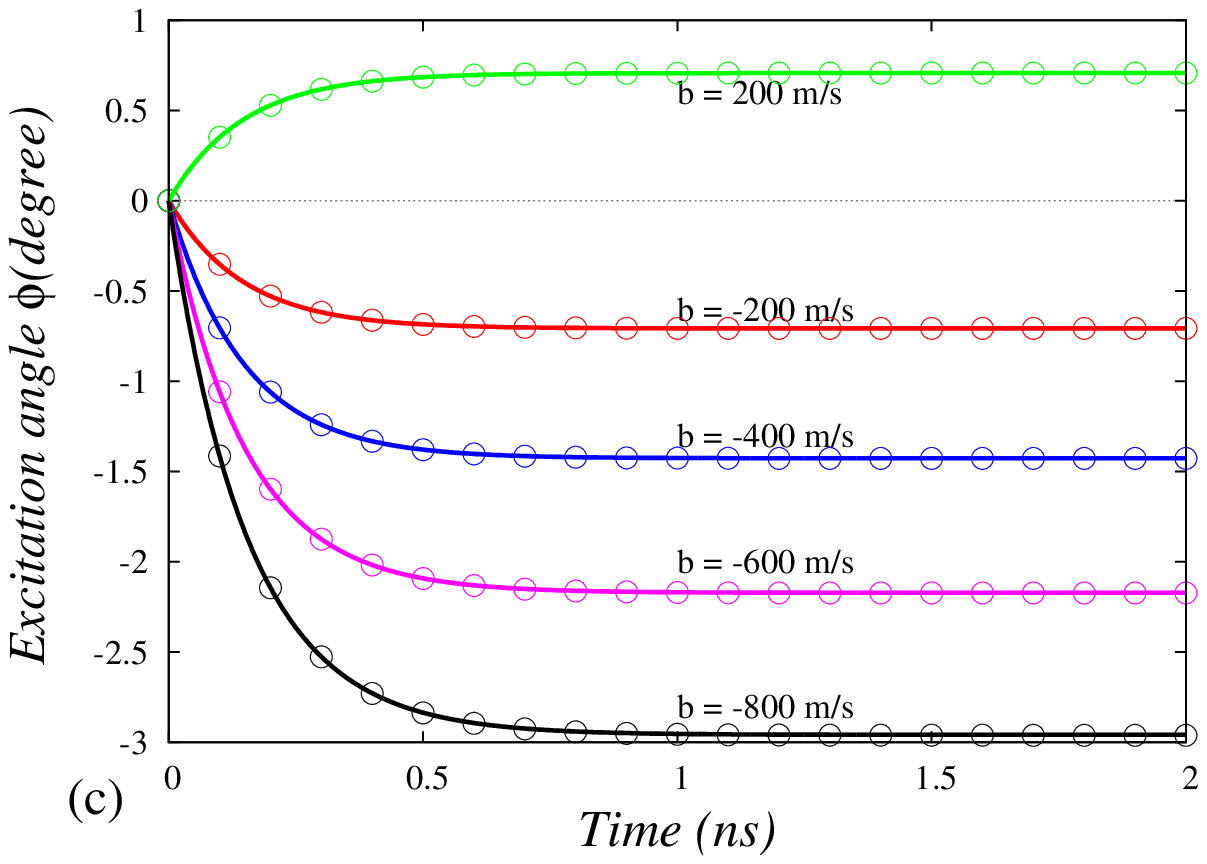}~\includegraphics[angle=0,width=0.5\linewidth]{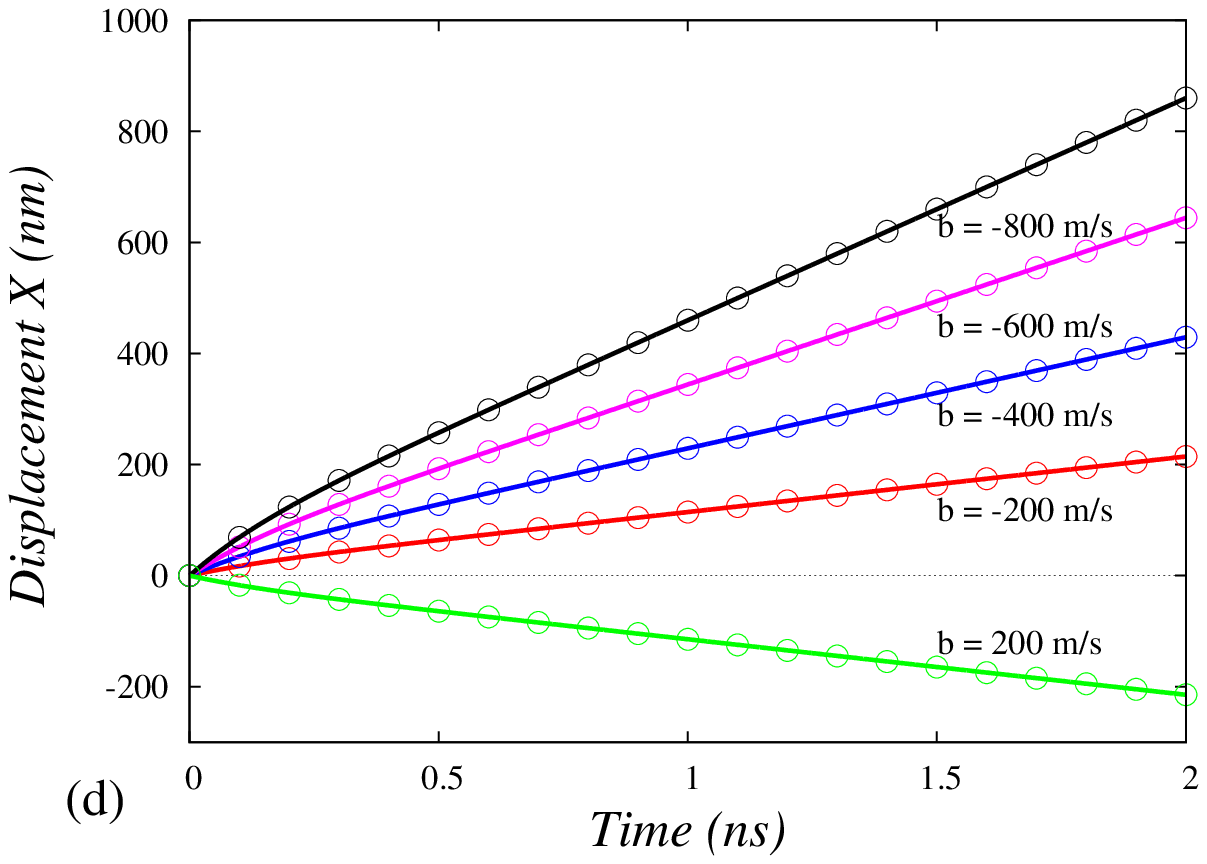}
\caption{(color online). The time variation of (a) velocity $v(t)$, (b) width $W(t)$, (c) excitation angle $\phi(t)$ and (d) displacement $X(t)$ of the Cobalt domain wall for different values of the current($b=$-200,-400,-600,-800 and 200 m/s) without driving field and transverse magnetic field.  The same colour of the solid line and open circle represent the analytical and numerical results respectively for the given spin-transfer torque.}
\end{figure}
\begin{figure}[!hbtp]
\centering\includegraphics[angle=0,width=0.8\linewidth]{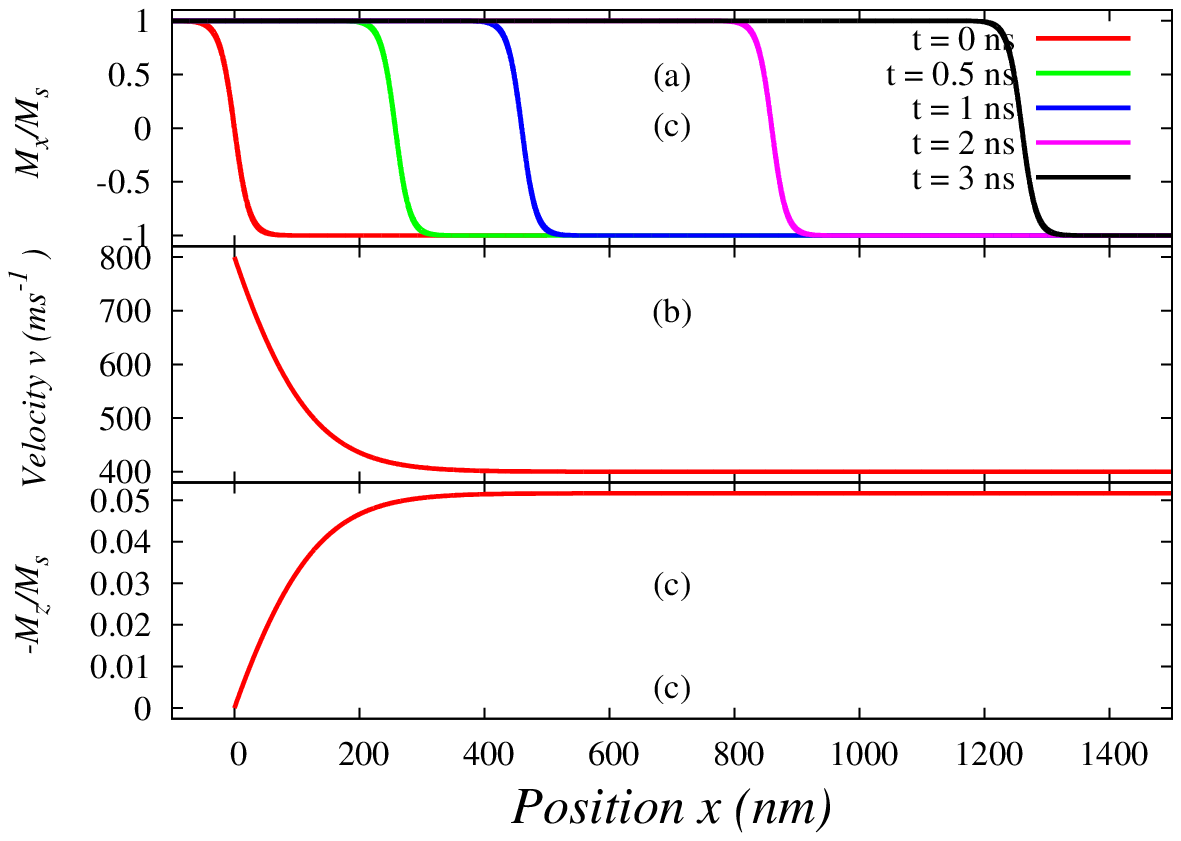}
\caption{(color online). (a) Spatial profile of the normalized x-component of magnetization $M_x/M_s$ of the wall for different time, (b) Velocity of the wall with respect to wall position and (c) Normalized z-component of the magnetization of the domain wall at the center of the wall as a function of wall position, for the value of adibatic spin-transfer torque $b=-800 m/s$ along with nonadiabatic spin transfer torque.}
\end{figure}
In order to substantiate the statement that the moving domain wall maintains the constant velocity and width, the analytical and numerical solutions for the velocity($v$), width ratio($W/W(0)$), excitation angle($\phi$) and displacement($X$) of the domain wall are plotted against time in FIGs.5(a-d) respectively for the different values of current($b$=-200,-400,-600,-800 and 200 m/s) in the absence of driving field and transverse magnetic field.   The open circle represents the numerical results and solid line represents the analytical results obtained from Eq.\eqref{dphi} and Eq.\eqref{phi} respectively. The quantities $\phi, v$ and $W/W(0)$ are initially changing with time and after a fraction of nanosecond they reach the saturated values such as $\phi_s,v_s,W_s/W(0)$ as shown in FIGs.5(a-c).  This is because of the equivalence between the rate of energy supplied by the incoming electrons and the rate of energy dissipation through damping.\cite{Zhang}

In FIG.5(a) we observe, initially the velocity of the domain wall seems to be equivalent to $-b$, and this can be explained by calculating the initial velocity $v(0)=-b(1+\alpha\xi)/(1+\alpha^2)\approx -b$ from Eq.\eqref{v} using the initial condition $\phi(0)=0$.  When the driving field and the transverse magnetic field are switched off, the saturated velocity of the domain wall is given by $v_s=-c/\alpha$ (Eq.\eqref{satvelocity1}). Hence, the adiabatic spin-transfer torque is considered to be most important for the initial velocity of the domain wall, whereas, the nonadiabatic spin-transfer torque controls the final velocity of the domain wall.

  Eventhough, the initial velocity is maximum and completely depends on the adiabatic spin-transfer torque($b$), $b$ itself cannot move the domain wall for a long distance, because the saturated velocity becomes zero without nonadiabatic spin-transfer torque.  The reason behind the vanishing of saturated velocity in the absence of nonadiabatic spin-transfer torque  is the absorbance of the adiabatic spin angular momentum of the incoming electrons by the distortion of the domain wall, caused by the adiabatic spin-transfer torque, so that the net adiabatic torque on the wall vanishes and as a consequence the wall stops.
The nonadiabatic spin-transfer toruqe($c$) is responsible for the nonzero final velocity of the domain wall, which behaves as a nonuniform magnetic field that can sustain a steady state motion of the wall.  Eventhough the magnitude of $c$ is about 2 orders smaller than the magnitude of $b$, the saturated velocity is large because it is inversely proportional to the damping parameter which is very small.  The reason for the initial velocity to be larger than the saturated velocity is due to the fact that the spin current supplies energy to the domain wall which is greater than the damping of the wall at the initial time.   As time increases, the damping increases through the distortion of the wall width and excitation of the domain wall.  When the supplied energy equals the damping energy, the wall moves at a constant velocity which is less than initial velocity\cite{Zhang1}.  It may be noted that, the direction of the velocity of the wall is always opposite to the direction of current, which means that the domain wall is dragged towards the direction of flow of electrons.
  Similar to the velocity, the width ratio of the domain wall also decreases with time for all values of current below the Walker limit and reaches the saturated value $W_s$ after a fraction of nanosecond.  However, when the current increases, the saturated value of width ratio decreases and the variation of the width is independent of the direction of the current as shown FIG.5(b).

The plot for the excitation angle $\phi(t)$ indicates that while the domain wall moves, the magnetic moments pointed in the positive x-direction precesses from -x to +x direction in a plane which is inclined by an angle $\phi$ with  positive y-direction is shown in FIG.5(c).  The change in the excitation angle indicates that the plane in which the precession of the magnetic moments of the domain wall takes place due to the adiabatic and nonadiabatic spin-transfer torques is forced out of the plane of the strip when the current is applied.  The excitation angle increases with time and reaches the saturated value $\phi_s$, which also increases with the current.  The saturation is attained due to the demagnetization field which tries to pull the magnetic moments from out of plane to inplane.  When $b>0$ and $b<0$ the magnetic moments of domain wall excite in the upward and downward direction to the plane of the strip respectively.  Using the numerical integration of Eq.\eqref{v}(open circle) and analytical solution of Eqs.\eqref{disp1}-\eqref{F}(solid line), the displacement of the domain wall is plotted in FIG.5(d).  Initially, the displacement of the domain wall is not linear with respect to time since the velocity is not a constant and later, the displacement becomes linear because of the velocity is constant.  And also it is observed that the domain wall is displaced in the direction opposite to the direction of the current.  From the plots corresponding to $b=$200 and -200 m/ in FIGs.5(a),(c) and (d), we observed that when the direction of current is reversed, the dynamical quantities velocity, the excitation angle and the displacement $X$ exhibit the same variation with time however with a reversed sign.

The spatial variation of the normalized x-component of magnetization $M_x/M_s$ at different times(t=0.0,0.5,1.0,2.0,3.0 ns) against the position of the center of the wall, to observe the displacement, velocity and distortion of the domain wall have been plotted in FIGs.6(a-c) respectively.  From FIG.6(a), it is observed that the wall is not stopped and moves with a constant velocity for a long distance due the nonadiabatic spin-transfer torque.  At $x=0$, the velocity of the wall is maximum and then decreases untill it reaches the saturated velocity and thereafter it remains constant as shown in FIG.6(b).  While the decrease in the velocity of the wall is due to the adiabatic spin angular momentum, its constancy is due to the nonadiabatic spin angular momentum\cite{Zhang}.  In FIG.6(c), the z-component of the normalized magnetization($M_z/M_s$) at the center of the domain wall has been plotted as a function of the position. The $M_z/M_s$ implies the excitation of the domain wall from the plane of the strip and it shows that the excitation of the domain wall increases initially and maintains as constant for a long distance.  Moreover, the saturation of excitation is very small, about $|M_z|/M_s \approx 0.06$, and one can understand that the domain wall is still of Neel wall type with small excitation.  
\subsection*{Case B: Effect of the transverse magnetic field on current and field driven domain wall motion($b,c,H_d,H_y\neq 0$)}
\begin{figure}[!hbtp]
\centering\includegraphics[angle=0,width=0.5\linewidth]{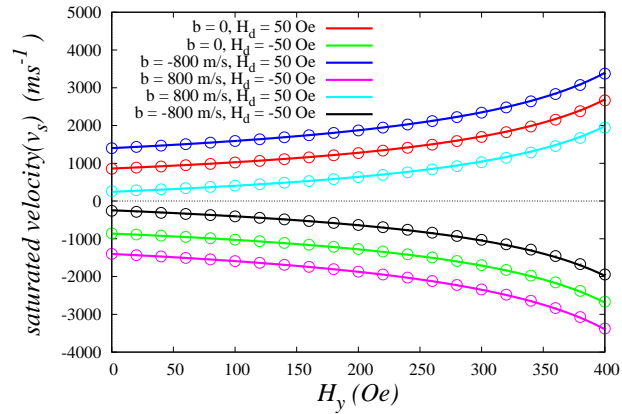}
\caption{(color online). The variation of the saturated velocity against the transverse magnetic field for different strengths of current and driving field in different directions have been plotted. The solid lines have been used for analytical results and the corresponding circles of the same color have been used for numerical results.}
\end{figure}
From the previous sections, it is observed that the initial velocity of the domain wall is controlled by the adiabatic spin-transfer torque and the final velocity is controlled by nonadiabatic spin-transfer torque and driving field. Further, we observed that the domain wall does not move and excite when the transverse magnetic field alone is applied. In this section, we study the impact of the transverse magnetic field on the saturated velocity of the domain wall driven by driving field and current and the results are plotted in FIG.7 between saturated velocity and transverse magnetic field for the different values of current and driving field. The solid line represents the analytical results which is plotted by using Eqs.\eqref{phi} and \eqref{satvelocity2} whereas the open circle indicates the numerical results which is obtained by solving Eqs.\eqref{dphi} and \eqref{satvelocity} numerically. 

The plots corresponding to $H_d=50~Oe,b=0~m/s$ and $H_d=-50~Oe,b=0~m/s$ show that the saturated velocity increases from 857~m/s  to 2666~m/s and -857~m/s  to -2666~m/s respectively when the transverse magnetic field is increased from 0 Oe to 400 Oe. Here, the saturated velocity is increased by the transverse magnetic field, because the magnetic energy of the domain wall decreases by the transverse magnetic field and the wall moves faster in order to have a larger rate of energy damping to dissipate the energy fastly.\cite{Boulle,Zhang1} The saturated velocity corresponding to the plots of  $b=-800~m/s,~H_d=50~Oe$ and $b = 800~m/s,~H_d=-50~Oe$ increases from 1400~m/s to 3374~m/s and -1400~m/s to -3374~m/s respectively when the transverse magnetic field is increased from 0 Oe to 400 Oe.
Similarly, the saturated velocity corresponding to the plots $b=800~m/s,~H_d=50~Oe$ and $b = -800~m/s,~H_d=-50~Oe$ increases from 259~m/s to 1944~m/s and -259~m/s to -1944~m/s respectively when the transverse magnetic field is increased from 0 Oe to 400 Oe.  These plots imply that the saturated velocity can be increased by the transverse magnetic field and it is irrespective of the directions of the driving field and current.  However, the saturated velocity is higher(lower) when the directions of driving field and current are antiparallel(parallel).  The reason for the suppression of saturated velocity when current and driving field are applied in the same direction, can be explained as follows.
In general, a domain wall moves in the opposite direction of current and same direction of driving field. Therefore, when both the current and the field are applied in the same direction, the motion by the current(driving field) will be opposed by the driving field(current) so that the velocity of the domain wall decreases.

In the above two cases(A and B), from the figures 5(a-d), 6(a-c) and 7, we can observe that the plots corresponding to the solid line resembles with the plots corresponding to the open circle.  This indicates that the obtained approximated analytical solutions given in Eqs.\eqref{phi} and \eqref{phi_zerocurrent} match with the numerical results.  Therefore, these analytical solutions can be used to understand the dynamics of domain wall in ferromagnetic nanostrip for the different materials.
\section{Conclusions}

In the present paper, the dynamics of transverse Neel wall in a ferromagnetic nanostrip in the presence of current, driving field and transverse magnetic field is studied by solving the Landau-Lifshitz-Gilbert equation with spin-transfer torques. By converting the LLG equation into spherical coordinates and using trial functions, the domain wall parameters such as the excitation angle, velocity, width and displacement are obtained. Under small angle approximation, the equation for $\phi$ is further reduced to the form of the Riccati equation and using its solution, the above mentioned four quantities in the presence of current, driving field and transverse magnetic field are obtained analytically. The results show that the initial velocity of the domain wall can be controlled by the adiabatic spin-transfer torque and the saturated velocity can be controlled by the nonadiabatic spin-transfer torque and driving field. The direction of the saturated velocity is antiparallel and parallel to the directions of the current and the driving field respectively. For the current driven domain wall motion with nonadiabatic spin-transfer torque, the saturated velocity of the wall linearly increases with the current.  But in the case of field driven domain wall motion, the saturated velocity no longer increases with the driving field, because the saturated velocity is proportional to the product of the driving field and the saturated width decreases with increase in the driving field. In the presence of a transverse magnetic field alone the domain wall is at rest, but the width of the wall is increased at the time of field which is applied and thereafter the width is also constant.

 Numerical results showed that the domain wall is driven by the current in the presence of the transverse magnetic field, the saturated velocity remains constant. However the initial velocity can be increased with transverse magnetic field when the initial value of the excitation angle $\phi$ is slightly perturbed.  Whereas for the field driven domain wall motion, the saturated velocity of the wall increases considerably, when the transverse magnetic field is increased. The transverse magnetic field is increased from 0 Oe to 400 Oe, the corresponding saturated velocity increases from 857 m/s to 2666 m/s, for the case of $H_d=50$ Oe, $b=0$ m/s and from 1400 m/s to 3374 m/s, for the case of $H_d=50$ Oe, $b=-800$ m/s.  Further, the numerical results showed that the saturated velocity is increased by the transverse magnetic field with the irrespective of the directions of the driving field and current and the saturated velocity is higher(lower) when the directions of driving field and current are antiparallel(parallel). The obtained approximated analytical solutions given in Eqs.\eqref{phi} and \eqref{phi_zerocurrent} match with the computed numerical results.

  In conclusion, the obtained analytical results of the dynamical parameters namely the excitation angle, velocity, width and displacement closely coincide with the numerical results.  While the transverse magnetic field has no effect on the saturated velocity of the domain wall in the current driven case, in the field driven case the saturated velocity of the domain wall increases by a large amount due to transverse magnetic field.  The compatibility of analytical results with numerical results implies that the analytical model with small angle approximation can be considered for transverse Neel domain wall dynamics in ferromagnetic nanostrips.
From the above, it is inferred that transverse magnetic field plays a crucial role in enhancing the saturated velocity of the domain wall which will be useful to design high and low speed domain walls depending upon the real time applications and also promises to make the efficient storage devices.

\section{Acknowledgement}
The work of M.D and R.A forms part of a major DST project. P.S wishes to thank DST Fast track scheme for providing financial support.  

\end{document}